\documentclass[final,3p,times]{elsarticle}

\usepackage{epsfig}
\usepackage{graphicx}
\usepackage{epstopdf}
\usepackage{color}
\usepackage{amssymb}
\usepackage{amsthm}
\journal{Annals of Physics}

\begin{document}

\begin{frontmatter}



\title{A multifunctional quantum thermal device: with and without inner coupling}

\author{Yong Huangfu }
\author{Shi-fan Qi}
\author{Jun Jing\corref{mycorrespondingauthor}}
\cortext[mycorrespondingauthor]{Corresponding author}
\ead{jingjun@zju.edu.cn}

\address{Department of Physics, Zhejiang University, Hangzhou 310027, Zhejiang, China}

\begin{abstract}
A three-level system attached to three thermal baths is manipulated to be a microscopic thermal device integrating a valve, a refrigerator, an amplifier, and a thermometer in the quantum regime, via tuning the inner coupling strength of the system and the temperatures of the external baths. We discuss the role of the inner coupling as well as the steady-state quantum coherence in these thermal functions using the Redfield master equation under a partial secular approximation. A high-sensitive thermometer for the low-temperature terminal can be established without the assistance from the inner coupling of the system. Our study of this multifunctional thermal device provides a deeper insight to the underlying quantum thermodynamics associated with the quantum coherence.
\end{abstract}

\begin{keyword}
\texttt
quantum thermodynamics\sep thermometer\sep inner coupling\sep quantum coherence
\end{keyword}

\end{frontmatter}




\section{Introduction}

Classical thermodynamics based on the macroscopic statistics has been developed over two hundred years forming a mature theoretical framework. However, it is always an interesting project to incorporate quantum mechanics into thermodynamics, which stems from a microscopic theory about discrete levels and quantum coherence. Many works have made an active exploration about the presence of quantumness in thermodynamics during the past two decades. These works on quantum thermodynamics can be roughly categorized into two divisions in terms of the quantum effect from environment~\cite{Scully-quantumness-2003,Abah-quantumness-2014,Guarnieri-quantumness-2018,Mukhopadhyay-quantumness-2018} and that from system~\cite{Kilgour-refrigerators-2018,Quan-quantumness-2007,Barrios-quantumness-2017,Dazhi-quantumness-2018}. The present work is devoted to investigate the steady-state coherence by the inner coupling of the system that is utilized as a multifunctional thermal device.

Various quantum thermodynamical functions, such as quantum valve~\cite{Lee-valve-2013,Tang-valve-2018,Yu-Valve-2019}, quantum refrigerator~\cite{Levy-Refrigerator-2012,Correa-refrigerators-2015,Segal-refrigerators-2018}, quantum amplifier~\cite{Li-amplifier-2012,Joulain-amplifier-2016}, quantum rectifier~\cite{Naseem-rectifier-2020}, quantum magnetometry~\cite{Bhattacharjee-magnetometry-2020} and quantum clock~\cite{Erker-clock-2017}, have raised growing attentions in recent years. These models provide not only fundamental platforms to test the macroscopic thermodynamic laws down to the level of quantum mechanics, but also valuable references to design the microscopic quantum devices with certain thermal functions by actively and precisely tune the heat flux via the external macroscopic parameters, such as temperatures.

The measurement on the temperatures, especially on the low-temperatures of a microscopic system, is an open problem in quantum thermodynamics. Precisely measuring the sample temperature in a very low region is a particularly important issue in the development of microscopic technologies. Thermometry at the microscopic scale has attracted a considerable amount of work in the fields of modern physics~\cite{Walker-TheMaterial-2003,kucsko-Thebio-2013,Zgirski-TheMaterial-2018,Mehboudi-Thequantum-2019,Mukherjee-thermometry-2019}. The smallest possible thermometer that composed merely of a single qubit was investigated in Ref.~\cite{Jevtic-Single-qubit-2015}. The authors studied the effect of initial-state quantum coherence on the thermometer to discriminate a cold bath from a hot one. But it is required for that thermometer to find out the temperatures of both baths in advance. Thus, the single-qubit thermometry determines whether a bath was hot or cold rather than exactly obtains the unknown temperature of the colder sample. Schemes of a practical thermometer in both theoretical and experimental aspects have been proposed in a variety of experimental platforms. For examples, optical thermometers were tested on the nitrogen-vacancy centers in diamond~\cite{kucsko-Thebio-2013,Neumann-TheNV-2013} and quantum-dot system~\cite{Haupt-Optical_Thermometry-2014,Seilmeier-Optical_Thermometry-2014}. Meanwhile, electronic thermometers have been available in the quantum-dot system~\cite{Rafael-Thedots-2011,zhang-Thedots-2019,Yang-Thedots-2019} and in the superconducting qubits~\cite{Hofer-Thermometer-2017}. In particular, the low-temperature thermometry proposed in Ref.~\cite{Hofer-Thermometer-2017} is performed by a quantum Otto engine coupled to a hot reservoir with a known temperature $T_h$ and a cold one with an unknown temperature $T_c$. Linear dependence of $T_c$ on $T_h$ was established when the heat engine is working with the Carnot efficiency. This result also implies to integrate multiple thermal functions into a single device.

We endeavor to study a multifunctional thermal device within the framework of the open-quantum-system theory. For a microscopic open system immersed in an environment, master equations allow to track the relevant degrees of freedom in both dynamics and the steady-state behavior, which are under the influence of all the other degrees of freedom that are not of the immediate interest or out of a microscopic control~\cite{Hofer-MEq-2017,Li-SSC-2014,huangfu-SSC-2018,Wang-metrology-2018,Wang-SSC-2019}. Particularly, a perturbative expansion with respect to the system-bath coupling strength is employed to obtain a master equation for the system part. During the ordinary derivation, a Markovian approximation is firstly applied to obtain a Redfield master equation, and then a further secular approximation leads to the master equation in the well-known Lindblad form~\cite{Hofer-MEq-2017,Breuer_book_Open-Quantum_2002}. The Redfield master equation retains the steady-state coherence of the system under a non-equilibrium environment~\cite{Li-SSC-2014,huangfu-SSC-2018,Wang-metrology-2018}, yet may loss the positivity of the reduced density matrix in the short-time scale. In contrast, the Lindblad master equation~\cite{Gorini-MEq-1976,Lindblad-MEq-1976} always gives rise to a valid reduced density matrix upon the full secular approximation, yet fails to track the system coherence in the long-time limit. It was recently reported that the positivity of the Redfield master equation can be recovered by a partial secular approximation~\cite{Farina-Psecular_approximation-2019,Cattaneo-Psecular_approximation-2019}.

In the present work, we investigate the steady-state quantum coherence induced by the inner coupling within a microscopic model that is comprised of  a three-level system and three external thermal baths with different temperatures. This generic model can be considered as a thermal device linking to three terminals. Non-vanishing quantum coherence in the steady state is totally determined by the inner coupling of the system and demonstrated in the steady state. Certain thermal functions, such as valve, refrigerator and amplifier can be realized in the quantum regime with the inner coupling. Based on the functions of valve and refrigerator, we further propose a quantum thermometer to measure the temperature of the coldest thermal bath. In the parametric space of the atomic-level spacings and the terminal temperatures, a sensitive thermometer can be realized at the working points of a quantum valve (also the onset of a quantum refrigerator).

The rest of this work is organized as following. In Sec.~\ref{model}, we present the full Hamiltonian of the model, including the interaction between the system and environments (the three thermal baths). Then we introduce the Redfield master equation under a partial secular approximation for this model in the presence and in the absence of inner coupling. In Sec.~\ref{SST-coh}, it is demonstrated that the quantum coherence in the steady state is induced by the inner coupling. In Sec.~\ref{T-MutiF}, we investigate the heat flux associated with thermal functions. Multiple thermal functions are consequently realized, such as the valve and the amplifier to control the target heat currents and the refrigerator to cool down the coldest bath. Also a theoretical scheme of a quantum thermometer for the coldest bath is presented without the assistance from the inner-coupling of the system and then simulated by a double quantum-dot system. In Sec.~\ref{Con}, we summarize the whole work.

\section{Model and methods}\label{model}

\begin{figure}
  \centering
  \includegraphics[width=0.5\textwidth]{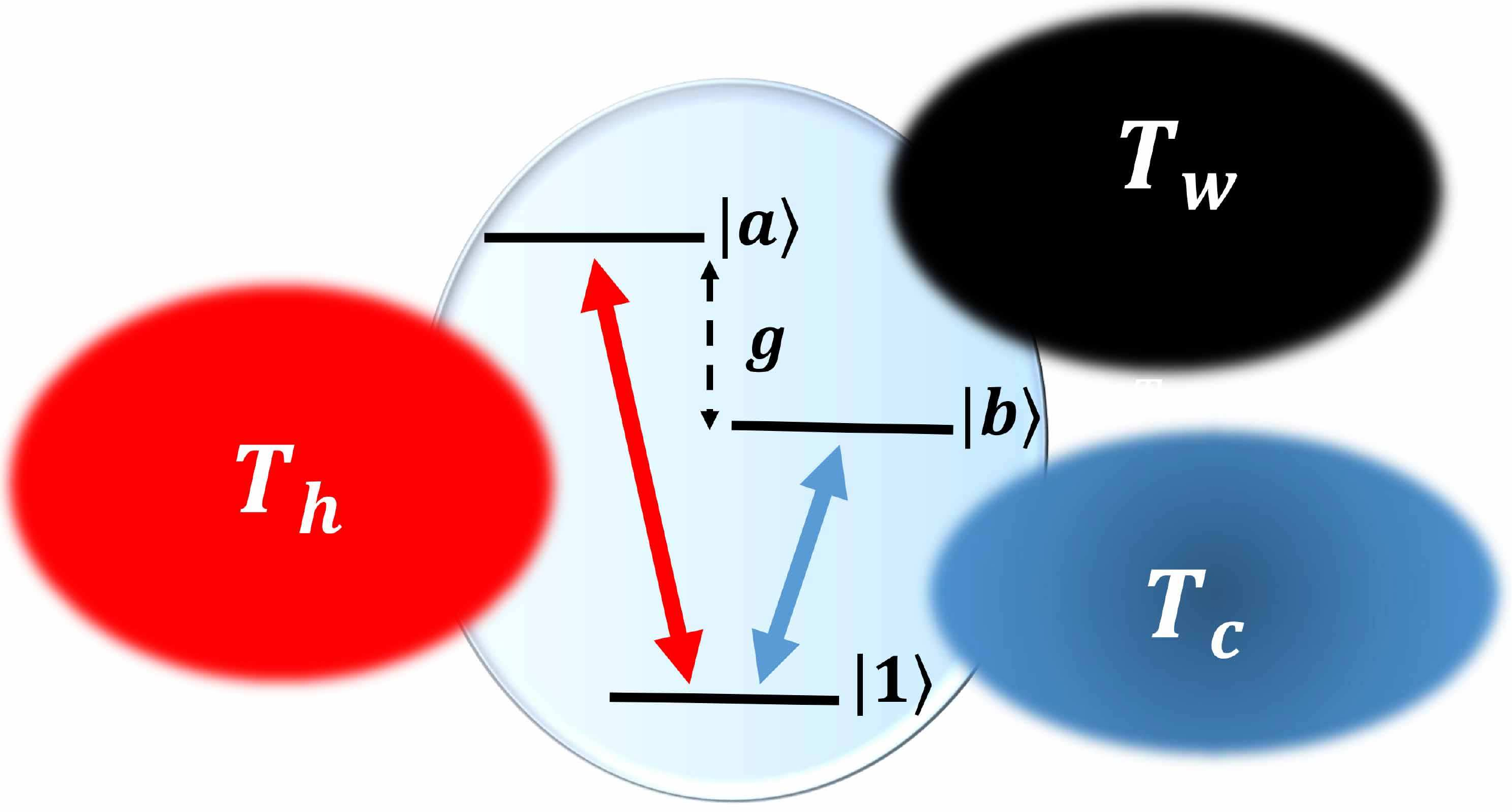}
  \caption{Diagram of a three-level system coupled to three terminals, which can be regarded as three baths with different temperatures labelled as the hot ($h$), cold ($c$), and work ($w$) bath, respectively.}\label{model_three}
\end{figure}

In a typical thermodynamics model demonstrated in Fig.~\ref{model_three}, a three-level microscopic system consisting of level-$1$, $b$ and $a$ by the increasing order of energy, is coupled to three terminals in their respective thermal equilibrium states. They are labelled respectively by $h$, $c$, and $w$ implying their functions and temperatures. The full Hamiltonian for this model can be decomposed into the system Hamiltonian, the bath Hamiltonian and the interaction Hamiltonian:
\begin{eqnarray}\label{Htot}
 H &=&  H_{S} + H_{B} +H_{SB}= H_{S} + \sum_{\mu=h,c,w} H_{\mu}+\sum_{\mu=h,c,w}S_{\mu}\otimes B_{\mu}.
\end{eqnarray}

In the system energy basis (also named the bare basis), the Hamiltonian can be written as ($\hbar\equiv1$)
\begin{equation}\label{HS}
H_S=\omega_{a}|a\rangle\langle a|+\omega_{b}|b\rangle\langle b|+g\left(|a\rangle\langle b|+|b\rangle\langle a|\right),
\end{equation}
where the ground energy is set as $\omega_1=0$ and $g$ is the inner coupling strength between the two excited states~\cite{Yudin-coherently_coupled-2005}. The degenerate situation for $\omega_{a}=\omega_{b}$ has been investigated in Ref.~\cite{Kilgour-refrigerators-2018} about the application of a quantum refrigerator. It can be covered by the present model. Each thermal bath is assumed to be a collection of decoupled harmonic oscillators with bosonic creation and annihilation operators $b^\dagger_{q,\mu}$ and $b_{q,\mu}$ for the mode $q$ with frequency $\omega_{q,\mu}$ in the bath $\mu$:
\begin{equation}
H_{\mu}=\sum_{q}\omega_{q,\mu}b^\dagger_{q,\mu}b_{q,\mu}.
\end{equation}
The collective bath operators $B_{\mu}$'s in Eq.~(\ref{Htot}) describe the instantaneous displacement of bath oscillators from the equilibrium state,
\begin{equation}
  B_\mu=\sum_q\lambda_{q,\mu}\left(b^\dagger_{q,\mu}+b_{q,\mu}\right).
\end{equation}
The system operators $S_{\mu}$ describe the interaction between the three-level system and the environments:
\begin{equation}\label{Sohc}
  S_{h}= |1\rangle \langle a| + |a\rangle \langle 1|,\,
  S_{c}= |1\rangle \langle b| + |b\rangle \langle 1|,\,
  S_{w}=(|a \rangle\langle a|-|b\rangle\langle b|)\sin\phi +(|a\rangle\langle b|+|b\rangle\langle a|)\cos\phi,
\end{equation}
where $\phi\equiv\arctan(2g/\Delta)$ with $\Delta\equiv\omega_{a}-\omega_{b}$. The operator $S_{w}$ depends explicitly on the inner coupling-strength $g$ of the system. Energy flux transfer among the hot, cold and work baths through the microscopic system.

\subsection{The three-level system with inner coupling}

The derivation onset of a microscopic master equation~\cite{Breuer_book_Open-Quantum_2002} for the open-quantum-system dynamics requires all the system operators to be expressed in eigenbasis $|\lambda\rangle$ of the system Hamiltonian. For the present model, the Hamiltonian~(\ref{HS}) of the three-level system with the inner coupling should be diagonalized as
\begin{eqnarray*}
&& H_S=\omega_{2}|2\rangle\langle2|+\omega_{3}|3\rangle\langle3|,
 \quad  \omega_{2,3}=\frac{1}{2}\left(\omega_{a}+\omega_{b}\mp\sqrt{4g^2+\Delta^2}\right),\\
&&  |2\rangle=\cos\frac{\phi}{2}|a\rangle-\sin\frac{\phi}{2}|b\rangle, \quad
  |3\rangle=\sin\frac{\phi}{2}|a\rangle+\cos\frac{\phi}{2}|b\rangle.
\end{eqnarray*}
Consequently, the coupling operators $S_{\mu}$, $\mu=h,c,w$, in the interaction Hamiltonian $H_{SB}$ are rewritten as
\begin{equation}
  S_{h}= \left(\cos\frac{\phi}{2}|1\rangle\langle2|+\sin\frac{\phi}{2}|1\rangle\langle 3|\right)+h.c.,\,
  S_{c}= \left(\cos\frac{\phi}{2}|1\rangle\langle3|-\sin\frac{\phi}{2}|1\rangle\langle 2|\right)+h.c., \,
  S_{w}= |2\rangle\langle3|+|3\rangle\langle2|.
\end{equation}
The operators $S_{h}$ and $S_{c}$ in the eigenbasis indicate that the two excited states $|3\rangle$ and $|2\rangle$ are simultaneously coupled to the ground state $|1\rangle$ through both the hot and the cold baths. These interaction channels as shown in Fig.~\ref{current1} would induce quantum coherence in the density matrix spanned by $\{|1\rangle,|2\rangle,|3\rangle\}$. The bath-$w$ does work to the system through the operator $S_{w}$, which has been exploited to the thermal rectification and the heat amplification~\cite{Wang-SSC-2019}.

As derived in~\ref{partial-SSe}, the Redfield master equation with a partial secular approximation about the present model can be explicitly expressed by
\begin{eqnarray}
&&  \dot{\rho}_{S}=-i[H_{S},\rho_{S}] +\sum_{\mu=h,c,w}\mathcal{D}_{\mu}[\rho_{S}]\label{ME1}, \\
&&\mathcal{D}_{h}[\rho_{S}]= \nonumber \Gamma^{+}_{h3}(\omega_{3})\mathcal{L}_{\tau_{13}}(\rho_{S})+\Gamma^{-}_{h3}(\omega_{3})\mathcal{L}_{\tau_{31}}(\rho_{S})+ \Gamma^{+}_{h2}(\omega_{2})\mathcal{L}_{\tau_{12}}(\rho_{S})+\Gamma^{-}_{h2}(\omega_{2})\mathcal{L}_{\tau_{21}}(\rho_{S})\\
&+&\nonumber \sum_{m=2,3}\bigg\{\Gamma^{-}_{h1}(\omega_{m})(\tau_{m1}\rho_{S}\tau_{1\bar{m}}+\tau_{\bar{m}1}\rho_{S}\tau_{1m})
+\Gamma^{+}_{h1}(\omega_{m})([\tau_{1m}\rho_{S},\tau_{\bar{m}1}]+[\tau_{1\bar{m}},\rho_{S}\tau_{m1}])\bigg\},\\
&& \mathcal{D}_{c}[\rho_{S}]=\nonumber \Gamma^{+}_{c2}(\omega_{3})\mathcal{L}_{\tau_{13}}(\rho_{S})+\Gamma^{-}_{c2}(\omega_{3})\mathcal{L}_{\tau_{31}}(\rho_{S})+  \Gamma^{+}_{c3}(\omega_{2})\mathcal{L}_{\tau_{12}}(\rho_{S})+\Gamma^{-}_{c3}(\omega_{2})\mathcal{L}_{\tau_{21}}(\rho_{S})\\   &-&\nonumber  \sum_{m=2,3}\bigg\{\Gamma^{-}_{c1}(\omega_{m})(\tau_{m1}\rho_{S}\tau_{1\bar{m}}+\tau_{\bar{m}1}\rho_{S}\tau_{1m})
+ \Gamma^{+}_{c1}(\omega_{m})([\tau_{1m}\rho_{S},\tau_{\bar{m}1}]+[\tau_{1\bar{m}},\rho_{S}\tau_{m1}])\bigg\},\\
&&\mathcal{D}_{w}[\rho_{S}]=\nonumber
   \Gamma^{+}_{w}(\Omega)\mathcal{L}_{\tau_{23}}(\rho_{S})+\Gamma^{-}_{w}(\Omega)\mathcal{L}_{\tau_{32}}(\rho_{S}),
\end{eqnarray}
where the Lindblad superoperator is defined as $\mathcal{L}_{X}(\rho_{S})\equiv2X\rho_{S} X^{\dag}-X^{\dag}X\rho_{S}-\rho_{S}X^{\dag}X$, with $X$ an arbitrary system operator $\tau_{nm}\equiv|n\rangle\langle m|$ ($\bar{m}\equiv5-m$) and $\Omega\equiv\omega_{3}-\omega_{2}=\sqrt{4g^2+\Delta^2}$. The transition rates can be factored as $\Gamma^{\pm}_{\mu l}(\omega)=f_{l}\Gamma^{\pm}_{\mu}(\omega)$, $l=1,2,3$ and $\mu=h,c,w$, where $f_{1}=\sin\frac{\phi}{2}\cos\frac{\phi}{2}$, $f_{2}=\cos^{2}\frac{\phi}{2}$, and $f_{3}=\sin^{2}\frac{\phi}{2}$. And $\Gamma^{\pm}_{\mu}(\omega)$ can be further factored as the product of $G_{\mu}(\omega)$ and $n_{\mu}(\omega)+1$ or $n_{\mu}(\omega)$, where $G_{\mu}(\omega)\equiv2\pi\sum_{q}|\lambda_{q,\mu}|^2\delta(\omega_{q,\mu}-\omega)$ is the spectral density function of bath-$\mu$ and $n_{\mu}(\omega)=(e^{\beta_{\mu}\omega}-1)^{-1}$ with $\beta_{\mu}=1/T_{\mu}$ ($k_B\equiv1$) is the average population determined by the temperature of bath-$\mu$. In this work, the three baths are assumed to be of Ohmic spectrum, i.e., $G_{\mu}(\omega)=\gamma_{\mu}\omega e^{-\omega/\omega_c}$, where $\gamma_{\mu}$ is a dimensionless system-bath coupling strength and $\omega_{c}$ is the cutoff frequency characterizing the largest energy scale.

Equation~(\ref{ME1}) consists of the unitary evolution term $-i[H_{S},\rho_{S}]$ and the dissipative term $\mathcal{D}_{\mu}[\rho_{S}]$, $\mu=h,c,w$, from the three baths. The first lines in both $\mathcal{D}_{h}[\rho_{S}]$ and $\mathcal{D}_{c}[\rho_{S}]$ are contributed to the conventional rotating-wave terms, where the superoperator $\mathcal{L}_{\tau_{ij}}(\rho_{S})$ in the standard Lindblad form describes the transition process from state $|j\rangle$ to state $|i\rangle$. While the second lines in them are the slowly-oscillating terms retained by the partial secular approximation, which indicate the quantum interference between two interaction channels. More derivation details about the partial-secular Redfield master equation as well as the transition/decay rates can be found in~\ref{partial-SSe}.

\subsection{The three-level system without inner coupling}

In this subsection, we assume a vanishing inner-coupling strength within the system, i.e., $g=0$. In this condition, the system Hamiltonian is diagonal in the bare basis $\{|1\rangle, |a\rangle, |b\rangle\}$. The coupling operators $S_{h}$ and $S_{c}$ remain the same as those in Eq.~(\ref{Sohc}), and the operator $S_{w}$ reduces to
\begin{equation}
S_{w}=|a\rangle\langle b|+|b\rangle\langle a|.
\end{equation}
Now the interaction with the work bath merely induces the energy exchange between the two excited states without affecting their energy levels.

The absence of the inner-coupling $g=0$ renders $\phi=0$. Thus the master equation~(\ref{ME1}) reduces to
\begin{eqnarray}
&&\dot{\rho}_{S} =-i[H_{S},\rho_{S}]+\sum_{\mu=h,c,w}\mathcal{D}_{\mu}[\rho_{S}]\label{ME2}, \\
&&   \mathcal{D}_{h}[\rho_{S}]\nonumber= \Gamma^{+}_{h}(\omega_{a})\mathcal{L}_{\tau_{1a}}(\rho_{S})+\Gamma^{-}_{h}(\omega_{a})\mathcal{L}_{\tau_{a1}}(\rho_{S}), \\
&&   \mathcal{D}_{c}[\rho_{S}] = \Gamma^{+}_{c}(\omega_{b})\mathcal{L}_{\tau_{1b}}(\rho_{S})+\Gamma^{-}_{c}(\omega_{b})\mathcal{L}_{\tau_{b1}}(\rho_{S}), \\
&&   \mathcal{D}_{w}[\rho_{S}]\nonumber=
   \Gamma^{+}_{w}(\Delta)\mathcal{L}_{\tau_{ba}}(\rho_{S})+\Gamma^{-}_{w}(\Delta)\mathcal{L}_{\tau_{ab}}(\rho_{S}).
\end{eqnarray}

Due to the absence of inner coupling within the two excited states, the interference between two interaction channels naturally ceases. Thus, we can employ the full secular approximation to obtain the Lindblad-form master equation~(\ref{ME2}). In the following, we will discuss the steady-state and thermal functions in the presence and absence of inner coupling separately.

\section{The steady state with and without quantum coherence}\label{SST-coh}

\begin{figure}[htpb]
\centering
\includegraphics[width=0.5\textwidth]{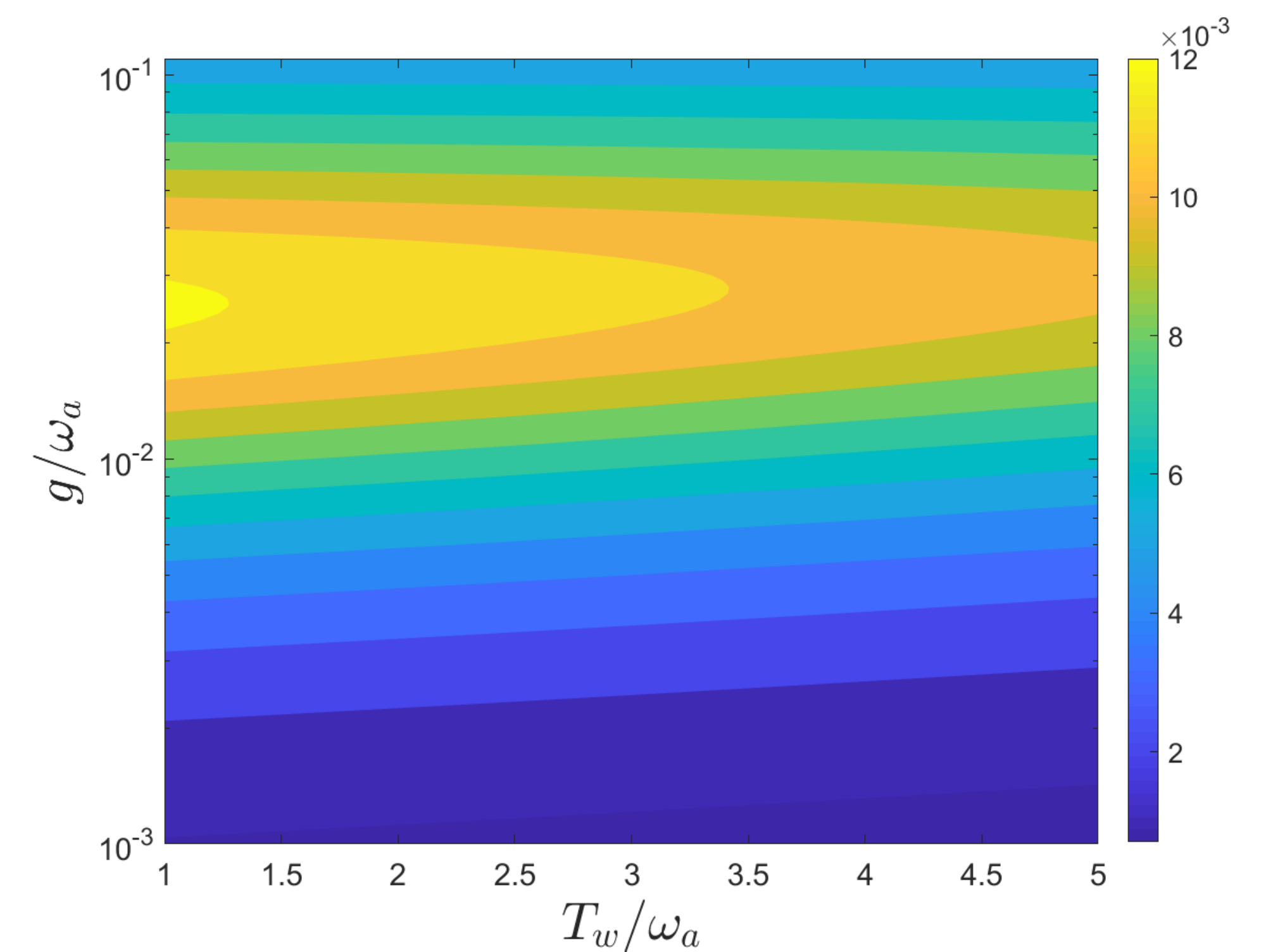}
\caption{(Color online) The steady-state coherence $|\rho_{23}|$ versus the inner coupling strength $g$ and the temperature of the work bath $T_{w}$. The other parameters are $\omega_{b}/\omega_{a}=0.95$, $\gamma_{h,c,w}/\omega_{a}=0.008$, $\omega_{c}/\omega_{a}=50$, $T_{h}/\omega_{a}=1$, and $T_{c}/\omega_{a}=0.1$.} \label{SSrho23}
\end{figure}

We have obtained the master equation beyond the full secular approximation by taking the slowly-oscillating terms into account. A direct consequence is to identify the quantum coherence in the steady state. The master equation~(\ref{ME1}) renders a steady state $\rho_{SS}$ in the following form:
\begin{equation}\label{rhoss1}
  \rho_{SS}=\left(\begin{array}{ccc} \rho_{11} & 0 & 0 \\  0 & \rho_{22} & \rho_{23} \\  0 & \rho_{32} & \rho_{33} \\  \end{array} \right),
\end{equation}
where the non-vanishing off-diagonal elements are determined by letting $\dot{\rho}_S=0$ in Eq.~(\ref{SSequations}). In Fig.~\ref{SSrho23}, the absolute value of coherence $|\rho_{23}|$ measuring the quantumness of the three-level system demonstrates a nontrivial dependence on the inner-coupling strength $g$ and the temperature of the work bath $T_w$. Remarkable steady-state coherence $|\rho_{23}|$ turns out with a moderate magnitude of $g$ and roughly decreases with an increasing $T_w$. It is shown that either a too low or a too high $T_w$ is not favourable to the residue quantumness of the three-level system under the three thermal baths after a long-time evolution. With the selected parameters in Fig.~\ref{SSrho23}, the maximal value of $|\rho_{23}|$ is achieved when $g/\omega_{a}=0.02$.

The coherence of steady state, however, does not exist with the lack of internal coupling. In the long-time limit, the steady-state solution derived from the master equation~(\ref{ME2}) renders a diagonal steady-state of the system:
\begin{equation}\label{rhoss2}
  \rho_{SS}=\left(
             \begin{array}{ccc}
              \rho_{11} & 0 & 0 \\
              0 & \rho_{bb} & 0 \\
              0 & 0 & \rho_{aa} \\
             \end{array}
           \right),
\end{equation}
where
\begin{eqnarray}
&&\nonumber\rho_{11} = \frac{\Gamma^{+}_{c}(\omega_{b})[\Gamma^{+}_{h}(\omega_{a})+\Gamma^{+}_{w}(\Delta)]+\Gamma^{+}_{h}(\omega_{a})\Gamma^{-}_{w}(\Delta)}{\Lambda}, \\   &&\nonumber\rho_{bb} = \frac{\Gamma^{-}_{c}(\omega_{b})[\Gamma^{+}_{h}(\omega_{a})+\Gamma^{+}_{w}(\Delta)]+\Gamma^{-}_{h}(\omega_{a})\Gamma^{+}_{w}(\Delta)}{\Lambda},  \\  &&\label{rhoss2so}
  \rho_{aa} = \frac{\Gamma^{-}_{h}(\omega_{a})[\Gamma^{+}_{c}(\omega_{b})+\Gamma^{-}_{w}(\Delta)]+\Gamma^{-}_{c}(\omega_{b})\Gamma^{-}_{w}(\Delta)}{\Lambda},
\end{eqnarray}
with the normalization factor  $\Lambda=\Gamma^{+}_{h}(\omega_{a})\Gamma^{+}_{c}(\omega_{b})+\Gamma^{+}_{c}(\omega_{b})\Gamma^{+}_{w}(\Delta)+\Gamma^{+}_{h}(\omega_{a})\Gamma^{-}_{w}(\Delta) +\Gamma^{+}_{h}(\omega_{a})\Gamma^{-}_{c}(\omega_{b})+\Gamma^{-}_{c}(\omega_{b})\Gamma^{+}_{w}(\Delta)+\Gamma^{-}_{h}(\omega_{a})\Gamma^{+}_{w}(\Delta)
+\Gamma^{-}_{h}(\omega_{a})\Gamma^{+}_{c}(\omega_{b})+\Gamma^{-}_{c}(\omega_{b})\Gamma^{-}_{w}(\Delta)+\Gamma^{-}_{h}(\omega_{a})\Gamma^{-}_{w}(\Delta)$.
The three quantum channels connecting the thermal baths and the system are separable in the absence of the inner coupling. The dynamics of the populations and the coherence in this model are therefore decoupled from each other. Consequently the quantum coherence vanishes in the long-time limit.

This steady-state solution in Eq.~(\ref{rhoss2so}) essentially satisfies the principle of detailed balance. Note the rates of $\Gamma^{+}_{\mu}(\omega)$ and $\Gamma^{-}_{\mu}(\omega)$ characterize the probabilities of the decay and excitation transitions, respectively. Due to the conservation of the total population in the steady state, the probabilities of the population gain must be equivalent to the population loss for each level. In particular, we have
\begin{eqnarray}
  &&\Gamma^{+}_{c}(\omega_{b})\rho_{bb}+\Gamma^{+}_{h}(\omega_{a})\rho_{aa} =\nonumber [\Gamma^{-}_{c}(\omega_{b})+\Gamma^{-}_{h}(\omega_{a})]\rho_{11},  \\
  &&\Gamma^{-}_{c}(\omega_{b})\rho_{11}+\Gamma^{+}_{w}(\Delta)\rho_{aa} = \nonumber [\Gamma^{-}_{w}(\Delta)+\Gamma^{+}_{c}(\omega_{b})]\rho_{bb}, \\
  &&\Gamma^{-}_{h}(\omega_{a})\rho_{11}+\Gamma^{-}_{w}(\Delta)\rho_{bb} = [\Gamma^{+}_{h}(\omega_{a})+\Gamma^{+}_{w}(\Delta)]\rho_{aa}. \label{DB}
\end{eqnarray}
The results in Eq.~(\ref{rhoss2so}) are then determined by the solution of Eq.~(\ref{DB}) and the normalization condition without invoking the master equation.

\section{ Thermal multifunctions}\label{T-MutiF}

\subsection{Thermal functions with inner coupling}\label{TFuncion}

A quantum system can be regarded as a microscopic thermal device when it is used to control the heat flux back and forth from the system to the baths. In this subsection, it is shown that the open three-level system with inner coupling can be utilized as an integrated multifunctional thermal device with functions as valve, refrigerator and amplifier within certain parametric regions.

\begin{figure}[htpb]
\centering
\includegraphics[width=0.5\textwidth]{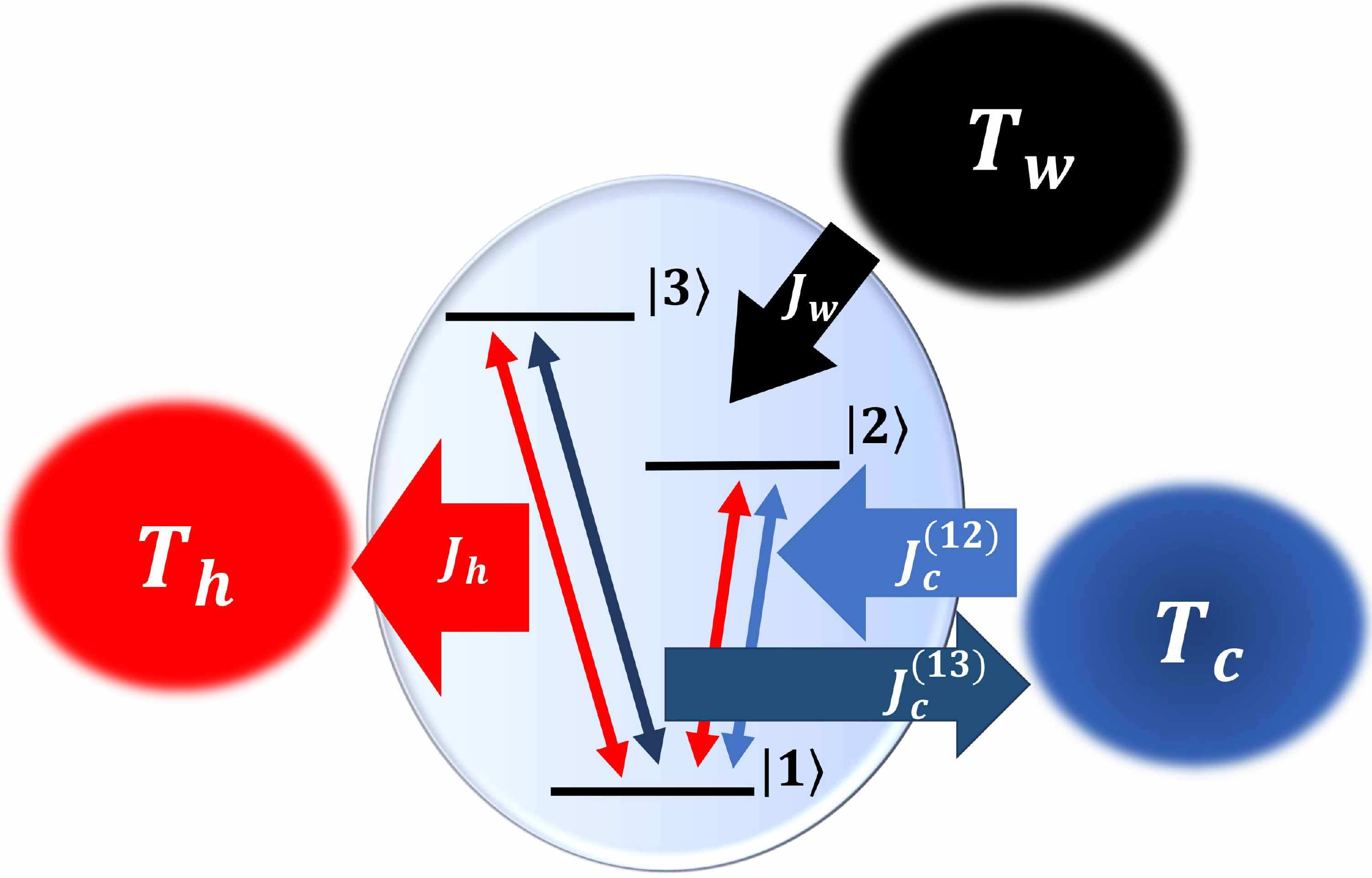}
\caption{(Color online) Diagram of the heat currents $J_{\mu}~(\mu=h,c,w)$ between the three-level system with inner coupling and the three thermal baths. $J_{c}$ is the heat current between the system and the cold bath, which is composed by $J^{(12)}_{c}$ and $J^{(13)}_{c}$ as displayed in Eq.~(\ref{Jqc1}).}\label{current1}
\end{figure}

The rate of the energy transfer between the system and bath-$\mu~(\mu=h,c,w)$, ${\rm Tr}[H_{S}\mathcal{D}_{\mu}(\rho_{S})]$, defines the corresponding heat current. In the long-time limit, $\rho_S$ takes the steady-state solution $\rho_{SS}$ given by Eq.~(\ref{rhoss1}). The standard formula for the (steady-state) rate of the heat exchange with bath-$\mu$~\cite{Kilgour-refrigerators-2018} is
\begin{equation}\label{Jq}
J_{\mu}\equiv{\rm Tr}\left[H_{S}\mathcal{D}_{\mu}(\rho_{SS})\right].
\end{equation}
Note a positive $J_{\mu}$ means the heat current follows the direction from bath-$\mu$ to the central system. By the master equation~(\ref{ME1}) and the current definition in Eq.~(\ref{Jq}), the steady-state energy current between the three-level system and the cold bath can be decomposed into two parts, each in charge of the coupling between the energy-level pair~$\{|1\rangle, |l\rangle\}$, $l=2,3$, and the cold bath:
\begin{equation}\label{Jqc1}
  J_{c}= \sum_{l=2,3}J^{(1l)}_{c}=\sum_{l=2,3}2\omega_{l}\big[\Gamma^{-}_{c\bar{l}}(\omega_{l})\rho_{11}-\Gamma^{+}_{c\bar{l}}(\omega_{l})\rho_{ll} +\Gamma^{+}_{c1}(\omega_{\bar{l}})(\rho_{23}+\rho_{32})\big],
\end{equation}
where $\bar{l}\equiv5-l$. Similarly, the heat current between the three-level system and the hot bath can be expressed by
\begin{equation}\label{Jqh1}
  J_{h}= \sum_{l=2,3}2\omega_{l}\big[\Gamma^{-}_{hl}(\omega_{l})\rho_{11}-\Gamma^{+}_{hl}(\omega_{l})\rho_{ll}- 2\Gamma^{+}_{h1}(\omega_{\bar{l}})(\rho_{23}+\rho_{32})\big].
\end{equation}
The steady-state quantum coherences, i.e., the terms $\rho_{23}$ and $\rho_{32}$, contribute to both $J_c$ and $J_h$. While the heat current between the three-level system and the work bath involves only with the populations on the two excited levels~$\{|2\rangle, |3\rangle\}$ of the system,
\begin{equation}\label{Jqw1}
J_{w}= 2\Omega\left[\Gamma^{-}_{w}(\Omega)\rho_{22}-\Gamma^{+}_{w}(\Omega)\rho_{33}\right].
\end{equation}
These currents in Eqs.~(\ref{Jqc1}), (\ref{Jqh1}) and (\ref{Jqw1}) are drafted in Fig.~\ref{current1}. They follow the law of energy conservation (the first law of thermodynamics): $J_{c}+J_{h}+J_{w}=0$. More detailed while cumbersome analytical expressions for the three currents can be directly obtained by the definitions of the decay rates $\Gamma_{\mu,l}^{\pm}$ and the matrix elements $\rho_{jk}$ of the steady state in Eq.~(\ref{rhoss1}). The following results about the steady-state currents under control are numerically obtained.

\begin{figure}[htpb]
\centering
\includegraphics[width=0.5\textwidth]{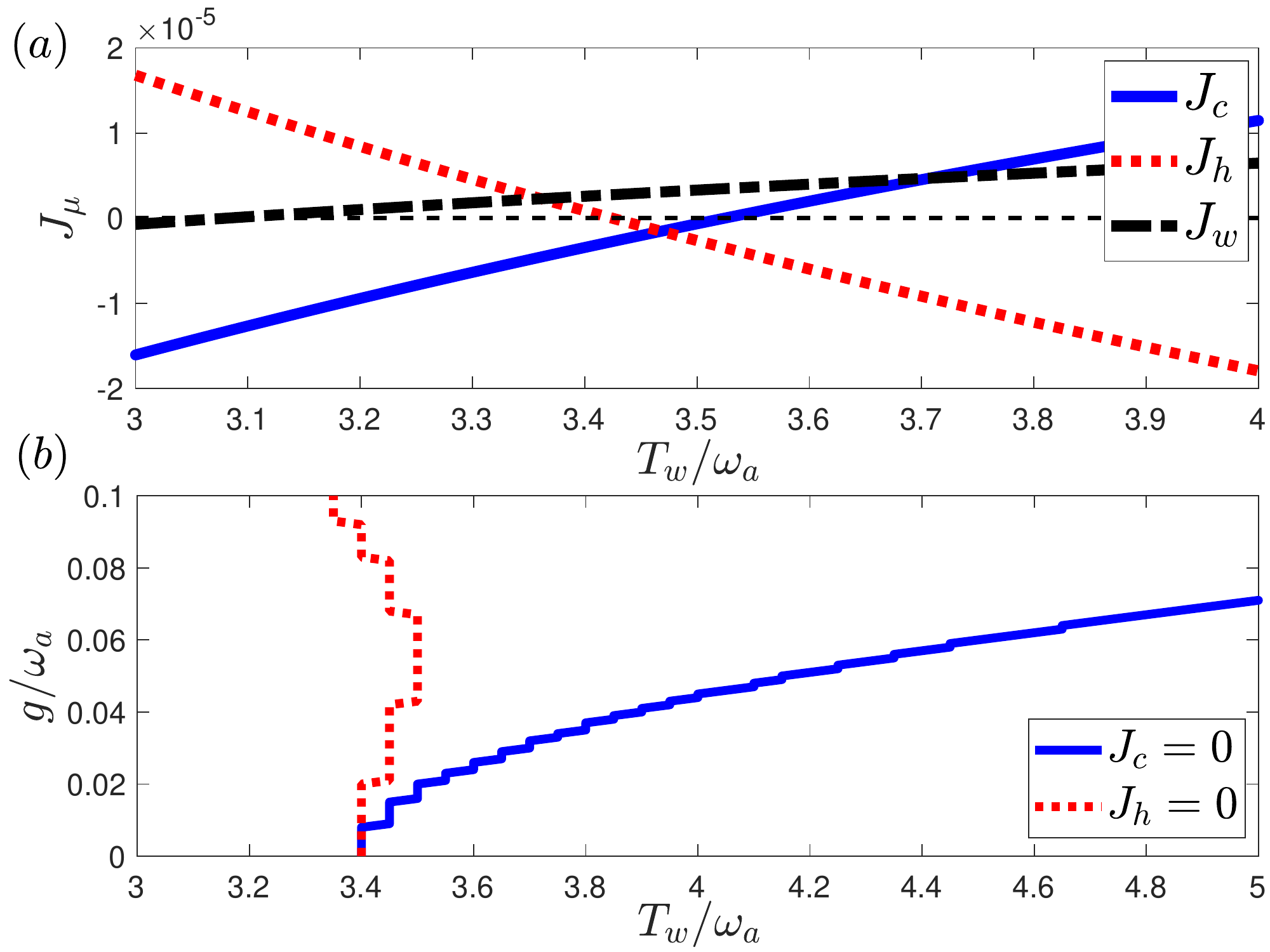}
\caption{(Color online) The three-level system in the steady-state can be regarded as a valve for the current $J_c$ or $J_h$. (a) The three heat currents $J_{\mu}~(\mu=c,h,w)$ versus the temperature of work bath $T_{w}$ with inner-coupling strength $g/\omega_{a}=0.02$. (b) The working spots ($J_h=0$ and $J_c=0$) of quantum valve by $T_{w}$ and $g$. The other parameters are set as $\omega_{b}/\omega_{a}=0.8$, $\gamma_{h,c,w}/\omega_{a}=0.008$, $\omega_{c}/\omega_{a}=50$, $T_{h}/\omega_{a}=1$ and $T_{c}/\omega_{a}=0.85$.}
\label{ther-Valve}
\end{figure}

Here the work bath is used as a control terminal to manipulate the three currents embodying various thermal functions in the quantum regime. In analogy to a classical valve, a quantum thermal valve can precisely cut off the heat current from any one of the terminals while leaving the currents to flow through the other two terminals under special conditions. The three heat currents with respect to the temperature of control terminal $T_{w}$ are demonstrated in Fig.~\ref{ther-Valve}(a), where the inner coupling strength $g$ is set as $0.02\omega_{a}$. When $T_{w}/\omega_{a}$ approaches a critical value, which is about $3.42$ with the parameters chosen in the plot, the energy current $J_h$ vanishes while the other two currents persist. Then when $T_{w}/\omega_{a}$ approaches the next critical value (about $3.53$), the heat current $J_c$ disappears. These critical values are fully determined by the settings of the system energy structure and the temperatures of the other two terminals. They are the working spots for the quantum valve, which can be manipulated by the inner-coupling strength $g$ as shown in Fig.~\ref{ther-Valve}(b). It is found that as long as $T_w$ is over a critical value, the quantum thermal valve for $J_c=0$ can be realized by increasing $g$. A higher $T_w$ corresponds to a larger $g$. In comparison, when $T_w$ has been properly determined by the other parameters, the choice of $g$ for $J_h=0$ is almost irrelevant to the value of $T_w$.

\begin{figure}[htpb]
\centering
\includegraphics[width=0.5\textwidth]{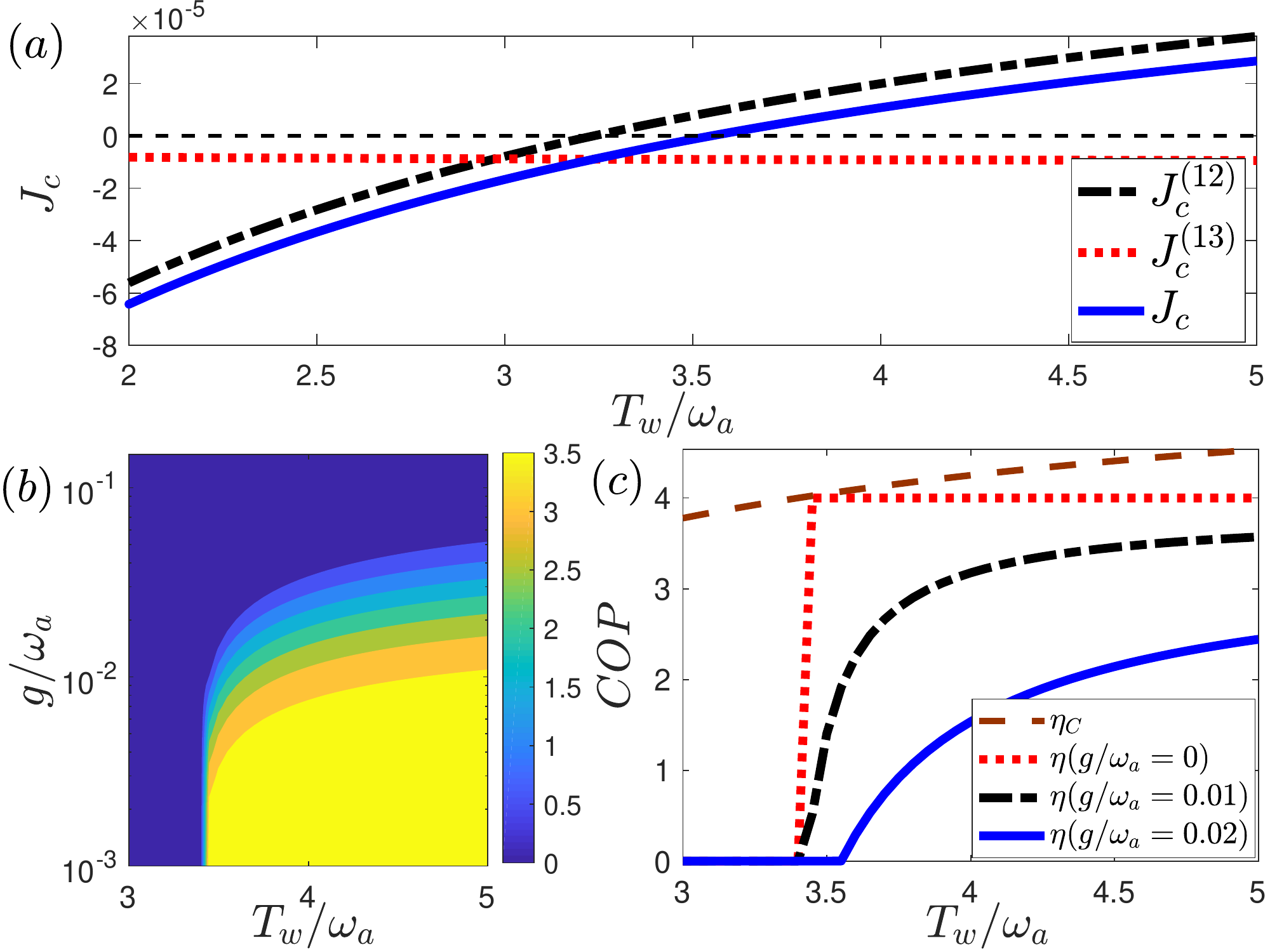}
\caption{(Color online) The three-level system under the steady state can also be regarded as a refrigerator for the coldest bath. (a) The current $J_{c}$ versus the temperature of work bath $T_{w}$ with inner-coupling strength $g/\omega_{a}=0.02$. (b) The coefficient of cooling performance (COP) under various $g$ and $T_{w}$. (c) COP versus $T_{w}$ with different inner coupling $g$. The other parameters are set as $\omega_{b}/\omega_{a}=0.8$, $\gamma_{h,c,w}/\omega_{a}=0.008$, $\omega_{c}/\omega_{a}=50$, $T_{h}/\omega_{a}=1$ and $T_{c}/\omega_{a}=0.85$. }\label{ther-Refri}
\end{figure}

Focusing on the direction of the heat flux between the system and the cold bath, one can find that our three-level system under the steady state can be used as a quantum refrigerator characterized by observing a net current out of the cold bath, i.e. $J_{c}>0$. It means that by controlling the working terminal, the system can continuously absorb the energy from the coldest terminal instead of importing energy to it. As shown in Fig.~\ref{ther-Refri}(a), the sign of $J_{c}$ switches from negative to positive when $T_{w}/\omega_{a}$ is over the critical point $3.53$ of the quantum valve. According to Eq.~(\ref{Jqc1}), $J_{c}$ can be decomposed into $J_c^{(12)}$ and $J_c^{(13)}$ (also shown in Fig.~\ref{current1}). The two sub-currents result from the coupling between the respective energy-level pairs of the system and the cold bath. One can see that the heat current $J^{(13)}_{c}$ stemming from the top level $|3\rangle$ is always negative, i.e., the top level $|3\rangle$ always conveys energy to the cold bath. It can be understood that the effective temperature for the energy-level pairs of $\{|1\rangle,|3\rangle\}$ is always higher than the temperature of bath-$c$. That is supported by the presence of inner coupling within the three-level system. In contrast, $J_c^{(12)}$ could become positive by increasing $T_w$. In the refrigerator regime depending on the working bath, $J_c>0$ and $J_w>0$. According to the first law of thermodynamics, the heat current $J_{h}$ follows the direction from the central three-level system to the bath-$h$, i.e., $J_{h}<0$. Figure~\ref{current1} describes all the directions of the heat currents between the refrigerator system and the three thermal baths.

The coefficient of cooling performance (COP) in the cooling regime, i.e., $J_{c}>0$, which is defined as $\eta\equiv J_{c}/J_{w}$~\cite{Kilgour-refrigerators-2018}, presents in Fig.~\ref{ther-Refri}(b). It is shown that the cooling efficiency of our quantum refrigerator will be gradually degraded with an increasing inner-coupling. In Fig.~\ref{ther-Refri}(c), we compare the COP of our refrigerator with the Carnot limit $\eta_{C}$:
\begin{equation}\label{etac2}
  \eta_{C}\equiv\frac{\beta_{h}-\beta_{w}}{\beta_{c}-\beta_{h}},
\end{equation}
which is the upper bound for the conventional thermal machines. Meanwhile, due to the Clausius theorem for the second law of thermodynamics, we have
\begin{equation}
\frac{\dot{Q}_{c}}{T_{c}}+\frac{\dot{Q}_{h}}{T_{h}}+\frac{\dot{Q}_{w}}{T_{w}}
=\frac{J_{c}}{T_{c}}+\frac{J_{h}}{T_{h}}+\frac{J_{w}}{T_{w}}\leq0,
\end{equation}
where $\dot{Q}_{\mu}$, $\mu=h,c,w$, is the heat flux between the system and bath-$\mu$. Based on the first and second laws of thermodynamics, one can obtain
\begin{eqnarray}\label{COPvsCarnot}
  \eta\equiv\frac{J_{c}}{J_{w}} \leq \frac{T_{h}-T_{w}}{T_{c}-T_{h}}\cdot\frac{T_{c}}{T_{w}}=\frac{\beta_{h}-\beta_{w}}{\beta_{c}-\beta_{h}}=\eta_{C}.
\end{eqnarray}
As displayed by the red dotted, the blue solid and the black dot-dashed lines in Fig.~\ref{ther-Refri}(c), $\eta$ is not greater than $\eta_{C}$ and becomes smaller for less $g$. At the onset point for the cooling window ($J_{c}=0$), the COP $\eta$ becomes equivalent to $\eta_C$ when $g=0$ as shown in the red dotted line. For a fixed $g$, $\eta$ is enhanced by increasing the working-bath temperature $T_w$. And it is found to be independent of $T_w$ when $g=0$. Further discussions about the situation without the inner coupling can be discovered in the next subsection [See around Eq.~(\ref{eta2})]. Since the presence of the inner coupling is closely correlated with the residue quantum coherence that demands extra work, one can conjecture that the cooling efficiency of the refrigerator is reduced by the quantum coherence. It is thus interesting to see that quantum coherence plays the role of quantum friction, which degrades the performance of microscopic heat engines~\cite{Francica-friction-2019,Roie-friction-2020}.

\begin{figure}[htpb]
\centering
\includegraphics[width=0.5\textwidth]{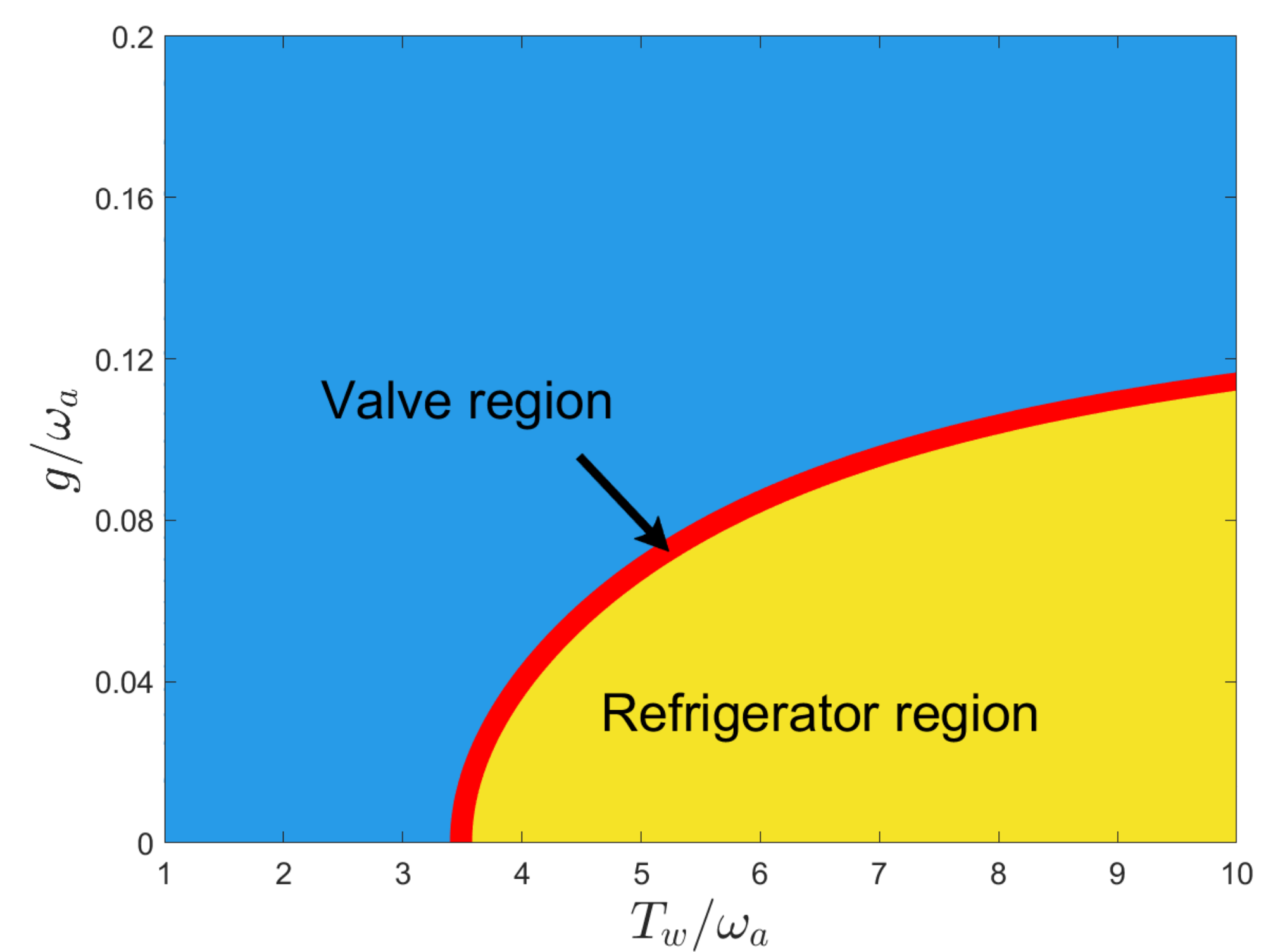}
\caption{(Color online) The phase diagram of the thermal functions of valve and refrigerator in the parametric space of the temperature of the work bath $T_{w}$ and the inner-coupling strength $g$. The other parameters are set as $\omega_{b}/\omega_{a}=0.8$, $\gamma_{h,c,w}/\omega_{a}=0.008$, $\omega_{c}/\omega_{a}=50$, $T_{h}/\omega_{a}=1$ and $T_{c}/\omega_{a}=0.85$. }\label{ther-VaRe}
\end{figure}

An overall picture for the thermal functions of valve and refrigerator can be illustrated by a phase diagram in Fig.~\ref{ther-VaRe} for the heat current $J_{c}$ with the temperature of the work bath $T_{w}$ and the inner coupling $g$. The red belt consisting of the working spots with $J_{c}=0$ for the quantum valve controlling the heat flux between the system and the cold bath separates the refrigerator regime (the yellow region) with $J_{c}>0$ from the remaining parametric regime. One can roughly observe that a strong inner coupling accompanied by a high-temperature work bath is useful to realize quantum valve and refrigerator.

\begin{figure}[htpb]
\centering
\includegraphics[width=0.5\textwidth]{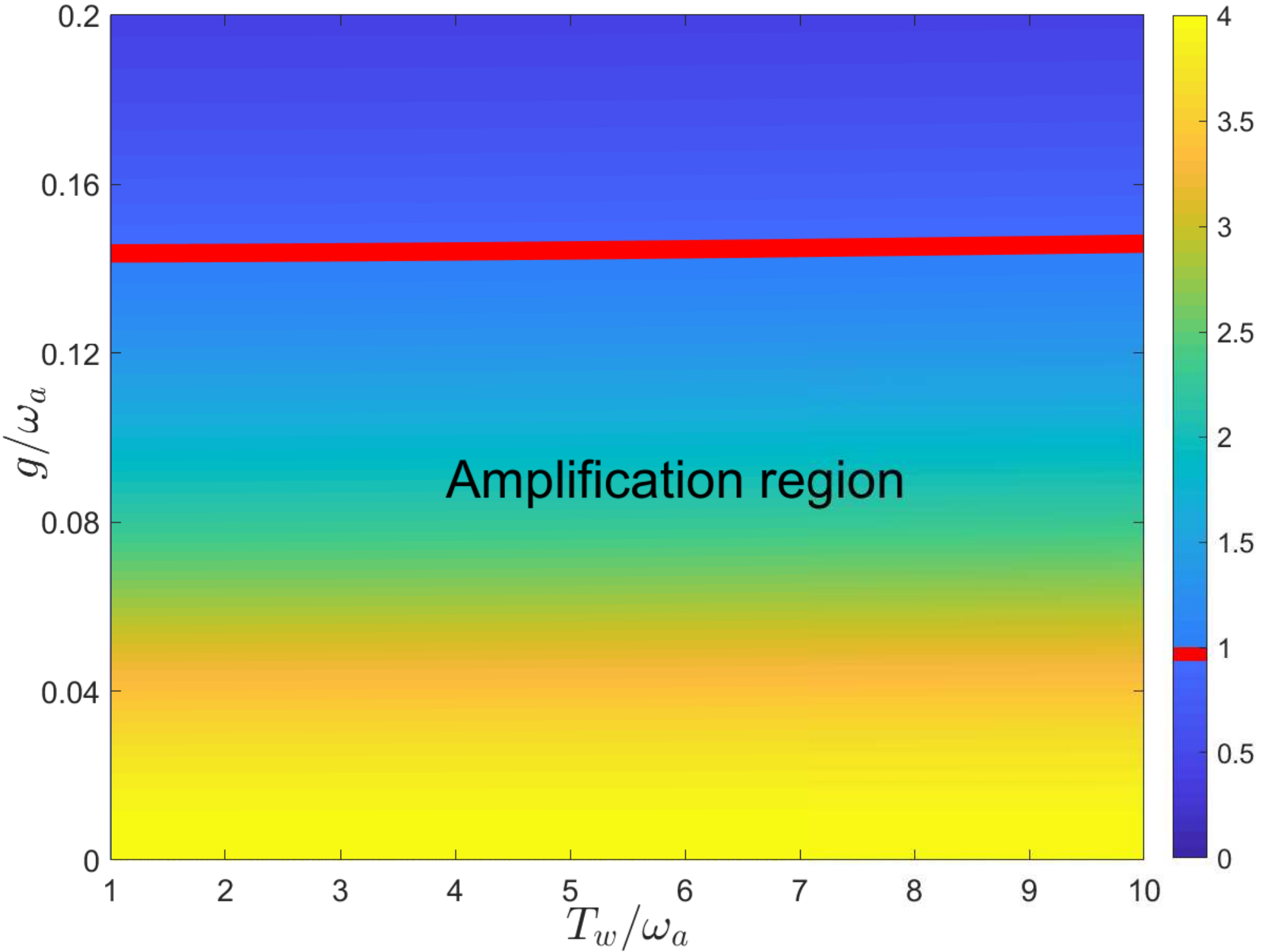}
\caption{(Color online) The amplification factor $\alpha_{J}$ under various temperatures of the work bath $T_{w}$ and the inner-coupling strength $g$. The other parameters are set as $\omega_{b}/\omega_{a}=0.8$, $\gamma_{h,c,w}/\omega_{a}=0.008$, $\omega_{c}/\omega_{a}=50$, $T_{h}/\omega_{a}=1$ and $T_{c}/\omega_{a}=0.85$.}\label{ther-Ampli}
\end{figure}

In the same parametric space as in Fig.~\ref{ther-VaRe}, we can also observe the region of a quantum thermal amplifier set up by our model. Here we focus on the variation of the heat flux between the work bath and the system $J_w$ upon that of the flux between the cold bath and the system $J_c$, i.e., the amplification factor~\cite{Li-amplifier-2012} $\alpha_{J}\equiv|\partial J_{c}/\partial J_{w}|$. A larger amplification factor $\alpha_J>1$ means a stronger amplification capacity about the working current $J_w$. The red line in Fig.~\ref{ther-Ampli} distinguishes the boundary between the amplification region $\alpha_{J}>1$ and the contraction region $\alpha_{J}<1$. It is clear that the microscopic three-level system with a small inner-coupling $g/\omega_a<0.142$ can be utilized as an amplifier. By the overlap between the region of valve and refrigerator in Fig.~\ref{ther-VaRe} and that of the amplifier in Fig.~\ref{ther-Ampli}, one can find that all of these thermal functions can be realized in our model with a finite inner-coupling strength $g$ under the same parametric setting.

\subsection{Microscopic Thermometer}\label{Thermometer}

While temperature is an intuitive notion deeply rooted in the daily life as well as the classical world, yet it is subtle and surprisingly difficult to formalise in quantum mechanics, especially in the field of low temperature region~\cite{Mehboudi-Thequantum-2019,Geusic-heat_pumps-1967}. In this subsection, we introduce a quantum thermometer to estimate the temperature of the coldest bath in our model of a three-level system without inner coupling. Here baths-$h$, $c$, and $w$ are respectively labelled as the conductor, sample, and control terminals in literature and it is assumed that the temperature of the sample terminal cannot be directly measured, in contrast to those of the conductor and the control terminals.

Due to the definition in Eq.~(\ref{Jq}), the steady-state energy current between the sample and the system is given by
\begin{equation}\label{Jqc2}
J_{c}=2\omega_{b}[\Gamma_{c}^{-}(\omega_{b})\rho_{11}-\Gamma_{c}^{+}(\omega_{b})\rho_{bb}].
\end{equation}
In comparison with Eq.~(\ref{Jqc1}), $J_{c}$ is now equivalent to $J^{(12)}_{c}$ having no contribution from quantum coherence. Similarly, the heat currents between the conductor and control terminals and the system are respectively given by
\begin{eqnarray}\label{Jqh2}
J_{h} &=& 2\omega_{a}[\Gamma_{h}^{-}(\omega_{a})\rho_{11}-\Gamma_{h}^{+}(\omega_{a})\rho_{aa}],\\
\label{Jqw2}J_{w} &=& 2\Delta[\Gamma_{w}^{-}(\Delta)\rho_{bb}-\Gamma_{w}^{+}(\Delta)\rho_{aa}].
\end{eqnarray}

\begin{figure}[htpb]
\centering
\includegraphics[width=0.5\textwidth]{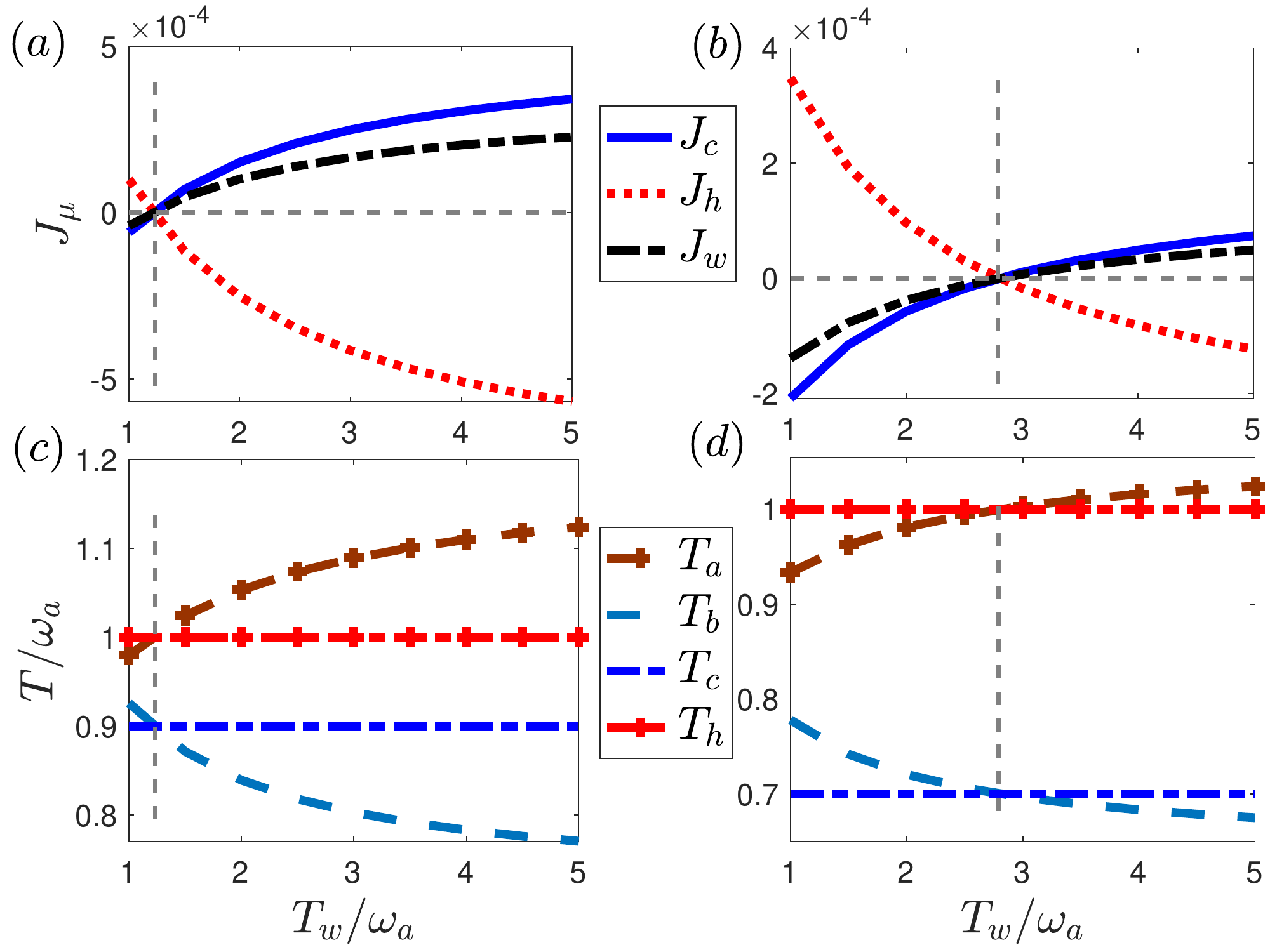}
\caption{(Color online) The thermal behaviors vary with the control temperature in the model without the inner coupling. (a) and (b): the heat currents towards the three terminals with $T_{c}/\omega_{a}=0.9$ and $T_{c}/\omega_{a}=0.7$, respectively. (c) and (d): the effective temperatures under the same conditions about $T_c$ in (a) and (b), respectively. The other parameters are set as $\omega_{b}/\omega_{a}=0.6$, $T_{h}/\omega_{a}=1$, $\gamma_{c,h,w}/\omega_{a}=0.008$ and $\omega_{c}/\omega_{a}=50$.} \label{The1}
\end{figure}

In Figs.~\ref{The1}(a) and (b), the heat currents are characterized by Eqs.~(\ref{Jqc2}), (\ref{Jqh2}), and (\ref{Jqw2}) with respect to the temperature of the control terminal $T_{w}$. It is shown that our three-level system in the absence of the inner coupling is a special thermal valve for both $J_c$ and $J_h$ when $J_w$ is switched off. The particular working points of the valve depend on the other parameters. The parameters containing the temperatures of the three baths would determine the population distributed on the three levels of our system. One can then define two effective temperatures for the energy-level pairs of $\{|1\rangle, |a\rangle\}$ and $\{|1\rangle, |b\rangle\}$ due to the Boltzmann distribution. They read,
\begin{equation}\label{effTem}
T_s=\omega_s/\ln\left(\frac{\rho_{ss}}{\rho_{11}}\right), \quad s=a, b.
\end{equation}
In general situations for the non-equilibrium steady state, the nonzero currents $J_c$ and $J_h$ will lead to $T_h\neq T_a$ and $T_c\neq T_b$ as demonstrated in Figs.~\ref{The1}(c) and \ref{The1}(d). Thus at the working points of quantum valve, for instances $T_w/\omega_a=1.2$ under $T_h/\omega_a=1$, $T_c/\omega_a=0.9$ in Fig.~\ref{The1}(c) and $T_w/\omega_a=2.8$ under $T_h/\omega_a=1$, $T_c/\omega_a=0.7$ in Fig.~\ref{The1}(d), all the heat currents vanish and the quantum system approaches a thermal equilibrium state. It thus guarantees $T_a=T_h$ and $T_b=T_c$ and the self-consistency about the effective temperatures. In another word, the populations of the three-level system in the thermal equilibrium state [see the two instances in Figs.~\ref{The1}(c) and \ref{The1}(d)] satisfy
\begin{equation}\label{pop}
\frac{\rho_{aa}}{\rho_{11}}=e^{-\frac{\omega_a}{T_h}}, \quad \frac{\rho_{aa}}{\rho_{bb}}=e^{-\frac{\Delta}{T_w}}, \quad
\frac{\rho_{bb}}{\rho_{11}}=e^{-\frac{\omega_b}{T_c}}.
\end{equation}
These expressions yield
\begin{equation}\label{pop2}
\frac{\omega_a}{T_h}=\frac{\Delta}{T_w}+\frac{\omega_b}{T_c}.
\end{equation}
Note again that the two $T_w$'s in Fig.~\ref{The1} are two specific cases in accordance to the choices of bath-temperatures, system energy configuration, and the coupling strength with the baths. In particular, a proper $T_w$ for an equilibrium state can always be obtained by sweeping over the parametric space when $T_c$ is fixed and it increases with a decreasing $T_c$. A more efficient cooling over a lower temperature sample requires more energy input from the control terminal with a higher temperature.

\begin{figure}[htpb]
\centering
\includegraphics[width=0.5\textwidth]{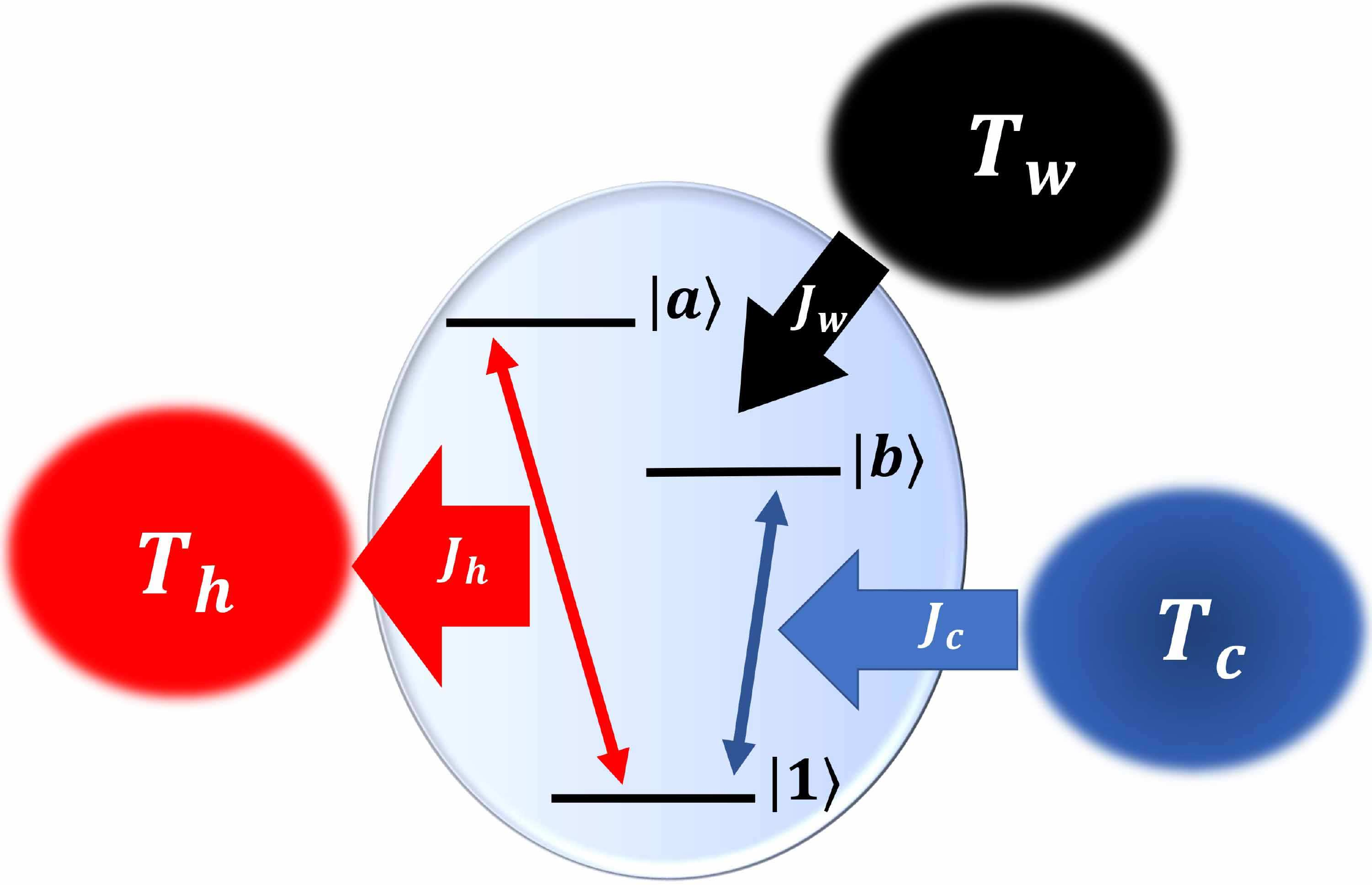}
\caption{(Color online) Diagram of the directions of heat flux $J_{\mu}~(\mu=h,c,w)$ between the system and the three thermal terminals in the absence of the inner coupling for the quantum refrigerator. } \label{current2}
\end{figure}

The three-level system under the thermal equilibrium state can be used to switch on the quantum functions of both valve and refrigerator. In particular, the working points ($J_{c}=0$) for the quantum valve are the onset points of cooling window for the quantum refrigerator ($J_{c}\geq0$). The coefficient of cooling performance of our system without the inner coupling is
\begin{equation}\label{eta2}
  \eta\equiv \frac{J_{c}}{J_{w}} =\frac{\omega_{b}}{\Delta},
\end{equation}
obtained by Eqs.~(\ref{DB}), (\ref{Jqc2}) and (\ref{Jqw2}). Comparing to the $T_w$-dependent result in the non-vanishing inner-coupling situation, now the COP $\eta$ becomes a constant $\omega_b/\Delta=\omega_b/(\omega_a-\omega_b)$ as the practical upper-bound shown in Fig.~\ref{ther-Refri}(c). Taking Eq.~(\ref{pop2}) into account, it is interesting to find that the equivalence in Eq.~(\ref{COPvsCarnot}), i.e., the Carnot limit, is achieved at the working points of the quantum valve. The Carnot efficiency can be attained in a quasi-static process. To reach the reversible limit, one needs in principle to work with infinitely slow cycles~\cite{Power-Esposito-2010}. The power of such a thermal device then becomes zero, which corresponds to the points with vanishing heat flux in our diagram. According to Eq.~(\ref{Jqc2}), the refrigerator condition $J_c\geq0$ means
\begin{equation}
\Gamma^{-}_{c}(\omega_{b})\Gamma^{+}_{h}(\omega_{a})\Gamma^{-}_{w}(\Delta)\geq
\Gamma^{+}_{c}(\omega_{b})\Gamma^{-}_{h}(\omega_{a})\Gamma^{+}_{w}(\Delta),
\end{equation}
where $\Gamma^{\pm}_{\mu}(\omega)=G_{\mu}(\omega)n_{\mu}(\mp\omega)$ with $G_{\mu}(\omega)$ the spectral density function and $n_{\mu}(\omega)$ the average population of bath-$\mu$. Immediately we have
\begin{equation}
  e^{\beta_{h}\omega_{a}}\geq e^{\beta_{c}\omega_{b}}e^{\beta_{w}\Delta},
\end{equation}
which is equivalent to Eq.~(\ref{COPvsCarnot}). Thus the onset of the cooling window as well as the refrigerator performance of our multifunctional device in the absence of the inner coupling is always consistent with the Carnot limit.

In the refrigerator regime, $J_c>0$ means $J_w>0$ and $J_h<0$. Their directions are shown in Fig.~\ref{current2}. One can therefore observe the heat current flows from the sample to the conductor through the three-level system. According to the definitions of the effective temperatures in Eq.~(\ref{effTem}), the effective temperatures of the system and the temperatures of terminal-$h$ and -$c$ are ordered by $T_{b}<T_{c}<T_{h}<T_{a}$.

More importantly, the sample temperature $T_c$ can be immediately obtained as
\begin{equation}\label{Tc-Tw}
T_c=\frac{T_{h}T_{w}\xi}{T_{w}-(1-\xi)T_{h}},
\end{equation}
where $\xi\equiv\omega_{b}/\omega_{a}$, by the thermal equilibrium condition. This relation between $T_c$ and the other two temperatures indicates an indirect measurement approach for $T_c$ by measuring the control temperature $T_w$ when the system approaches a thermal-equilibrium state. In addition, the energy-configuration parameter for the three-level system $\xi$ and the temperature of conductor terminal $T_h$ are supposed to be determined in advance.

\begin{figure}[htpb]
\centering
\includegraphics[width=0.5\textwidth]{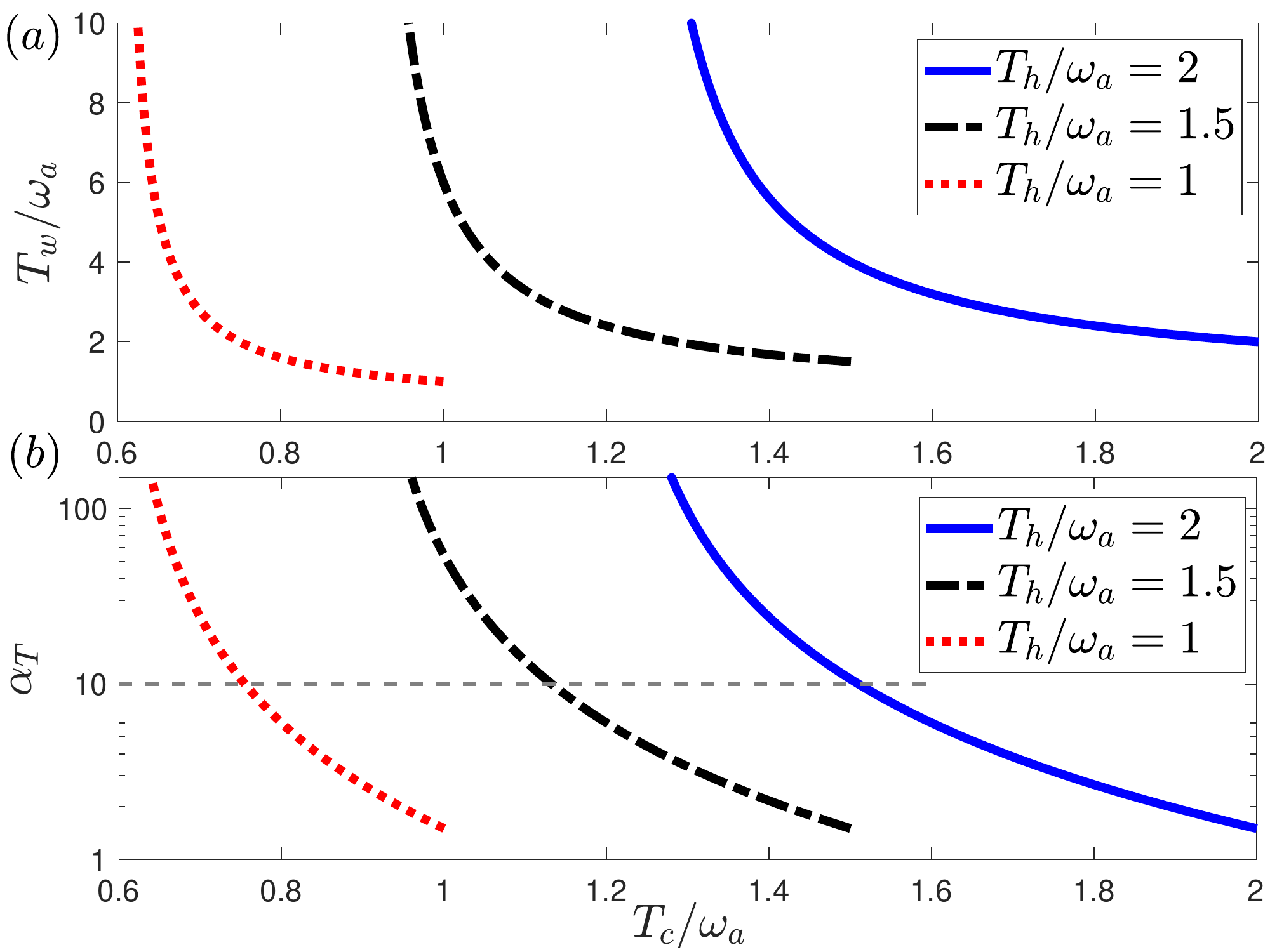}
\caption{(Color online) (a) the control temperature $T_w$ as a function of the sample temperature $T_c$; (b) the measurement sensitivity for the quantum thermometer $\alpha_T$ as a function of $T_c$ under different conductor temperature $T_h$. The energy-configuration parameter is set as $\xi=0.6$.} \label{The3}
\end{figure}

From Eq.~(\ref{Tc-Tw}), the lower bound of the sample temperature is determined by
\begin{equation}\label{lowTem}
  \lim_{T_{w}\rightarrow\infty}T_{c}= \xi T_{h}=\frac{\omega_{b}}{\omega_{a}}T_{h},
\end{equation}
meaning that the measurement range of our thermometer for $T_c$ is linearly proportional to both $T_h$ and the energy-configuration parameter $\xi$. This result under an infinite high-temperature control terminal coincides with that obtained by a previous proposal about the quantum thermometer associated with a quantum thermal machine~\cite{Hofer-Thermometer-2017}, in which the sample temperature $T_{c}$ has a linear relationship with the conduct terminal $T_{h}$. In the present thermometer scheme, however, we have a non-linear relation between the sample temperature and the control temperature:
\begin{equation}
  \frac{T_{c}}{T_{w}} = \frac{\xi}{\xi-1}+\frac{\xi}{(\xi-1)^2}\frac{T_{w}}{T_{h}}+\frac{\xi}{(\xi-1)^3} \left(\frac{T_{w}^2}{T_{h}^2}\right)+\mathcal{O}\left(\frac{T_{w}^3}{T_{h}^3}\right).
\end{equation}
Thus our proposal might provide a more sensitive measurement on $T_c$ than the existing one.

In Fig.~\ref{The3}(a), we display the dependence of the control temperature $T_w$ on the sample temperature $T_{c}$ under three different conductor temperatures $T_h$ and a fixed parameter $\xi=0.6$. Note $T_c\leq T_h$ according to the definition and $T_c\geq\xi T_h$ due to the constraint by Eq.~(\ref{lowTem}). As also shown in Fig.~\ref{The3}(a), the variation of $T_w$ with respect to $T_c$ is rapidly enhanced by decreasing $T_c$. This behavior is more clear in the measurement sensitivity~\cite{Mehboudi-Thequantum-2019} $\alpha_{T}$ shown in Fig.~\ref{The3}(b), which is defined as the absolute value of the control temperature bias divided by the sample-temperature variation:
\begin{equation}
  \alpha_{T}\equiv\left|\frac{\partial T_{w}}{\partial T_{c}}\right|=\frac{\xi(1-\xi)T^2_{h}}{(\xi T_{h}- T_{c})^2}.
\end{equation}
Then for a prescribed critical value of the sensitivity $\alpha_T$, a high-precision measurement region for the sample temperature are confirmed as $(\xi T_h, T'_{c}]$, where the so-called critical sample temperature $T'_{c}$ is
\begin{equation}\label{senTem}
  T'_{c}=\xi T_{h}+\sqrt{\frac{\xi(1-\xi)T^2_{h}}{\alpha_T}}.
\end{equation}
In Fig.~\ref{The3}(b), we set $\alpha_T=10$ as displayed by the horizontal dashed line. Then the cross points of the vertical lines and the horizontal line are critical temperatures $T'_{c}(\alpha_T)$. For example, when $T_{h}/\omega_{a}=2$ and $\xi=0.6$, one can perform a high-precision and indirect measurement for the sample temperatures in the range $1.2<T_{c}/\omega_{a}\leq1.51$ with at least one-order amplification in the magnitude of the temperature variation.

\subsection{Experimental simulation of the microscopic low-temperature thermometer}\label{Expri}

The quantum-dot systems have been widely applied in the field of quantum thermodynamics to simulate or realize various thermal functions, such as the heat engine~\cite{Rutten-ExHamiltonian-2009,Szukiewicz-three-terminal-2016,Jiang-three-terminal-2018}, the refrigerator~\cite{Prance-three-terminal-2009,Zhang-three-terminal-2015,Koski-three-terminal-2015, Erdman-three-terminal-2018,Dar-three-terminal-2019}, the novel energy carrier~\cite{Rafael-Thedots-2011,Walldorf-three-terminal-2017,Holger-three-terminal-2015}, and a Maxwell demon in the strong-coupling regime~\cite{Strasberg-three-terminal-2018}. Here we can show that a quantum-dot system can also be used to display the thermometer function in the proceeding subsection, by simulating the heat currents with the charge currents.

\begin{figure}[htpb]
\centering
\includegraphics[width=0.5\textwidth]{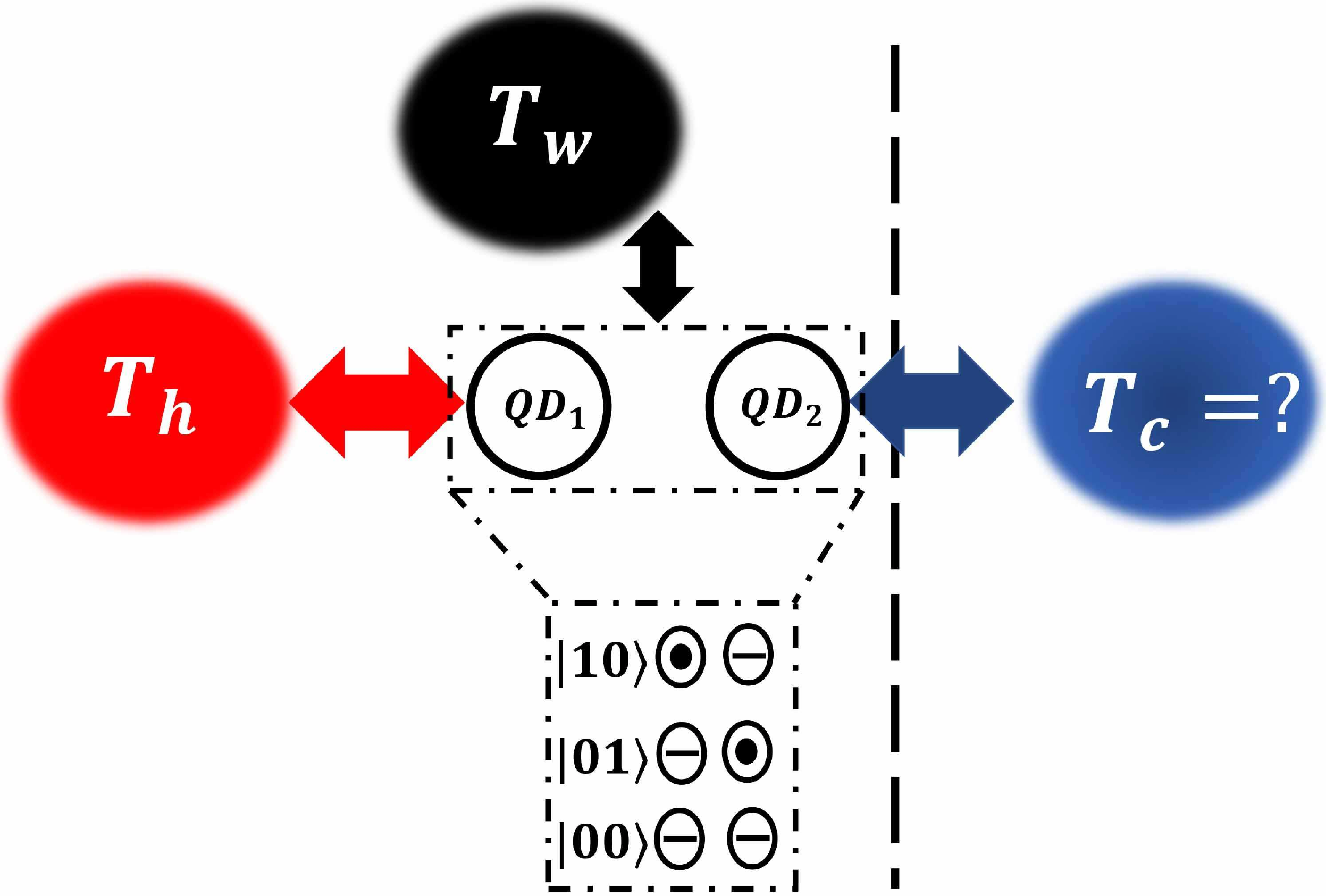}
\caption{(Color online) An experimental scheme for the quantum thermometer established in our microscopic device. It can be realized by an open double-quantum-dot system having a single electron coupled to three independent baths consisting of two metal leads and a radiation field. The three levels $|1\rangle$, $|b\rangle$, and $|a\rangle$ of the central system in our theoretical model (see Fig.~\ref{current2}) can be mapped to the electron states $|00\rangle$, $|01\rangle$, and $|10\rangle$, respectively.}\label{EX}
\end{figure}

A quantum-dot thermometer with three terminals is drafted in Fig.~\ref{EX}. The three-level system can be mimicked by a double-quantum-dot system which consists of two capacitively coupled quantum dots~(QD$_1$ and QD$_2$) operated in the Coulomb-blockade regime. The system is coupled to three terminals characterized by the two metal leads and a radiation field, which are labelled respectively by $h$, $c$, and $w$~\cite{Rutten-ExHamiltonian-2009}. The temperatures of the three terminals satisfy the same ordering assumed in our theoretical proposal: $T_{c}<T_{h}<T_{w}$. The terminal-$h(c)$ only exchanges electrons with the QD$_{1(2)}$. The corresponding Hamiltonian~\cite{Rutten-ExHamiltonian-2009} is
\begin{eqnarray}
  && H = H_{0}+V_{M}+V_{P}, \\
  && H_{0} = \sum_{m=1,2}\varepsilon_{m}c^{\dagger}_{m}c_{m} + \sum_{\mu=h,c}\varepsilon_{\mu}b^{\dagger}_{\mu}b_{\mu}+\sum_{\alpha}\omega_{\alpha}a^{\dagger}_{\alpha}a_{\alpha}, \label{ExH0}\\
  && V_{M} = (V_{1h}c_{1}b^{\dagger}_{h}+V_{2c}c_{2}b^{\dagger}_{c})+\mathrm{h.c.}, \label{ExVM}\\
  && V_{P} = \sum_{\alpha}(V_{\alpha}a_{\alpha}c^{\dagger}_{2}c_{1}+\mathrm{h.c.}). \label{ExVP}
\end{eqnarray}
The three terms in the free Hamiltonian $H_{0}$ of Eq.~(\ref{ExH0}) describe the isolated quantum dots, the free leads, and the radiation field, respectively. $c~(c^{\dagger})$, $b~(b^{\dagger})$ and $a~(a^{\dagger})$ are the annihilation (creation) operators for the electron in QD, the electron along the lead, and the photons, respectively. The Hamiltonian $V_{M}$ in Eq.~(\ref{ExVM}) describes the interaction between the quantum dots and the metal leads. The Hamiltonian $V_{P}$ in Eq.~(\ref{ExVP}) describes the dot-radiation-field interaction, addressing both the driving and the spontaneous light emission. Similar Hamiltonians can be also included in Refs.~\cite{Galperin-ExHamiltonian-2005,Galperin-ExHamiltonian-2006}.

A remarkable feature of the Coulomb-coupled quantum-dot system is that the electron transport through the system is forbidden but the capacitive coupling between the two dots allows electronic fluctuations to transmit heat between the terminals~\cite{Rafael-Thedots-2011,Zhang-Coulomb-coupled-2018}. Coulomb interactions prevent two electrons from being attended by the same time in the Coulomb-blockade regime. The Coulomb-coupled quantum-dot system in the subspace spanned by $\{|00\rangle, |01\rangle, |10\rangle\}$ constitutes an artificial three-level atom~\cite{Rafael-Thedots-2011,Rutten-ExHamiltonian-2009,Zhang-three-terminal-2015}, where the energy levels $|1\rangle$ and $|0\rangle$ indicate the state of the quantum dot occupied by one and zero electron, respectively. Then the three states of the center system in our model are mapped to the double-dot system in the single-electron subspace by $|1\rangle\leftrightarrow|00\rangle$, $|a\rangle\leftrightarrow|10\rangle$ and $|b\rangle\leftrightarrow|01\rangle$. Different dot energies $\varepsilon_{1}$ and $\varepsilon_{2}$ can be regarded as the energy gaps $\omega_a$ and $\omega_b$, respectively. The energy configuration of the quantum-dot system [$\xi$ in Eq.~(\ref{Tc-Tw})] can be determined by manipulating the gate voltage in advance~\cite{Prance-three-terminal-2009}.

The thermometer by the quantum-dot system can be implemented by the following procedure. As indicated in Fig.~\ref{The1}, one can start from $T_w=T_h$, and increase $T_w$ until $J_h=0$, which is in experiment the charge current between the central system and the conductor terminal. At this working-point, the quantum-dot system actually approaches a thermal equilibrium state. Then the temperature of sample terminal $T_c$ can be obtained by measuring $T_w$ according to Eq.~(\ref{Tc-Tw}).

\section{Conclusion}\label{Con}

In conclusion, we have investigated the steady-state thermal functions realized in an open-quantum-system model consisting of a three-level system attached to three thermal baths with or without inner coupling. The residue quantum coherence in steady-state is obtained by the Redfield master equation with a partial secular approximation. The residue quantum coherence in steady-state depends on the presence of inner coupling within the three-level system. It is found that a microscopic multifunctional thermal device can be established by varying the inner coupling within the system and the temperatures of the external baths. We identify the respective working regimes for the thermal functions as valve, refrigerator, and amplifier. It can be concluded that the coherence of steady-state may damage the cooling efficiency that is expected to achieve an optimal value.

We also demonstrate that the three-level system can be utilized as a microscopic thermometer to determine the temperature of the coldest terminal in the absence of the inner coupling. The quantum thermometer is established when the three-level system reaches the thermal equilibrium state with its three terminals, which also switches on the functions of valve and refrigerator. We find a non-linear dependence of the sample-temperature on the hot and the control temperatures, which could be exploited to perform a high-precision measurement. Our work therefore provides a deep insight to understand the roles of the system quantumness and the inner-coupling in quantum thermodynamics.

\section*{Declaration of competing interest}

The authors declare that they have no known competing financial interests or personal relationships that could have appeared to influence the work reported in this paper.

\section*{Acknowledgements}

We acknowledge grant support from the National Science Foundation of China (Grants No. 11974311 and No. U1801661), Zhejiang Provincial Natural Science Foundation of China under Grant No. LD18A040001, and the Fundamental Research Funds for the Central Universities (No. 2018QNA3004).

\appendix

\section{The master equation under the partial secular approximation}\label{partial-SSe}

In this appendix, we derive the master equation~(\ref{ME1}) in the main text under the partial secular approximation. It is of a Redfield master equation that conserves the positivity of the reduced density matrix even in the short-time scale.

The interaction Hamiltonian describing the coupling between the three-level system and the three thermal baths in Eq.~(\ref{Htot}) can be written as
\begin{equation}
H_{SB}=\sum_{\mu=h,c,w}H^{\mu}_{SB}=\sum_{\mu=h,c,w}S_{\mu}\otimes B_{\mu}.
\end{equation}
In the interaction picture with respect to $H_{S}+H_{B}$, it becomes
\begin{equation}\label{HII1}
H_{SB}(t)=\sum_{\mu=h,c,w}H^{\mu}_{SB}(t)=\sum_{\mu=h,c,w}S_{\mu}(t)\otimes B_{\mu}(t),
\end{equation}
where
\begin{eqnarray}
  S_{h}(t) &=& \cos\frac{\phi}{2}\left(\tau_{12}e^{-i\omega_{2}t}+\tau_{21}e^{i\omega_{2}t}\right) +\sin\frac{\phi}{2}(\tau_{13}e^{-i\omega_{3}t}+\tau_{31}e^{i\omega_{3}t}),\\
  S_{c}(t) &=& \cos\frac{\phi}{2}\left(\tau_{13}e^{-i\omega_{3}t}+\tau_{31}e^{i\omega_{3}t}\right) -\sin\frac{\phi}{2}(\tau_{12}e^{-i\omega_{2}t}+\tau_{21}e^{i\omega_{2}t}),\\
  S_{w}(t) &=& \tau_{23}e^{-i\Omega t}+\tau_{32}e^{i\Omega t},
\end{eqnarray}
with $\tau_{mn}\equiv|m\rangle\langle n|$ and $\Omega\equiv\omega_{3}-\omega_{2}$. The collective bath operators are
\begin{equation}
B_{\mu}(t)=\sum_{q}\lambda_{q,{\mu}}\left( b_{q,{\mu}}e^{-i\omega_{q,\mu}t}+b^\dagger_{q,{\mu}}e^{i\omega_{q,\mu}t}\right).
\end{equation}

Under the assumptions of a weak coupling between the microscopic system and the thermal baths (Born-approximation) and an ignorable relaxation time-scale for the thermal baths (Markovian approximation), one can apply the Redfield master equation to investigate the central-system dynamics to the second order of the coupling strength:
\begin{equation}\label{ME0}
\dot{\rho}_S=-\int_{0}^{\infty}ds{\rm Tr}_{B}[H_{SB}(t),[H_{SB}(t-s),\rho_{S}\otimes\rho_{B}]].
\end{equation}
The commutator operator in Eq.~(\ref{ME0}) expands as
\begin{eqnarray}
&&\nonumber [H_{SB}(t),[H_{SB}(t-s),\rho_{S}\otimes\rho_{B}]]=
 H_{SB}(t)H_{SB}(t-s)\rho_{S}\rho_{B}-H_{SB}(t)\rho_{S}\rho_{B}H_{SB}(t-s) \\
&-& H_{SB}(t-s)\rho_{S}\rho_{B}H_{SB}(t)+\rho_{S}\rho_{B}H_{SB}(t-s)H_{SB}(t).\label{HHPP}
\end{eqnarray}
Partial tracing over the degrees of freedom of the baths upon substituting the interaction Hamiltonian~(\ref{HII1}) into the master equation~(\ref{ME0}) turns out to be a summation over $12$ terms. The derivation is tedious but straightforward. One of the typical terms presents as following:
\begin{eqnarray}
&& {\rm Tr}_{B}\left[H^{h}_{SB}(t)\rho_{S}\rho_{B}H^{h}_{SB}(t-s)\right] \nonumber =\Gamma^{+}_{h3}(\omega_{3})[\tau_{13}\rho_{S}\tau_{13}e^{-2i\omega_{3}t}+\tau_{13}\rho_{S}\tau_{31}]+ \Gamma^{+}_{h1}(\omega_{3})[\tau_{13}\rho_{S}\tau_{12}e^{-i(\omega_{2}+\omega_{3})t}\\
&+&\nonumber\tau_{13}\rho_{S}\tau_{21}e^{-i(\omega_{3}-\omega_{2})t}]
+\Gamma^{-}_{h3}(\omega_{3})[\tau_{31}\rho_{S}\tau_{13}+\tau_{31}\rho_{S}\tau_{31}e^{2i\omega_{3}t}] + \Gamma^{-}_{h1}(\omega_{3})[\tau_{31}\rho_{S}\tau_{12}e^{i(\omega_{3}-\omega_{2})t}
+ \tau_{31}\rho_{S}\tau_{21}e^{i(\omega_{3}+\omega_{2})t}]\\&+&\nonumber
\Gamma^{+}_{h2}(\omega_{2})[\tau_{12}\rho_{S}\tau_{12}e^{-2i\omega_{2}t} +\tau_{12}\rho_{S}\tau_{21}]
+ \Gamma^{+}_{h1}(\omega_{2})[\tau_{12}\rho_{S}\tau_{13}e^{-i(\omega_{2}+\omega_{3})t}+\tau_{12}\rho_{S}\tau_{31}e^{-i(\omega_{2}-\omega_{3})t}]\\
&+&\nonumber
\Gamma^{-}_{h2}(\omega_{2})[\tau_{21}\rho_{S}\tau_{12}+\tau_{21}\rho_{S}\tau_{21}e^{2i\omega_{2}t}] +\Gamma^{-}_{h1}(\omega_{2})[\tau_{21}\rho_{S}\tau_{13}e^{i(\omega_{2}-\omega_{3})t}+\tau_{21}\rho_{S}\tau_{31}e^{i(\omega_{2}+\omega_{3})t}].\label{HpH1}
\end{eqnarray}
Here the decay rates are defined as $\Gamma^{\pm}_{h l}(\omega)=\Gamma^{\pm}_{h}(\omega)f_{l}$ with the factors $f_{1}=\sin\frac{\phi}{2}\cos\frac{\phi}{2}$, $f_{2}=\cos^{2}\frac{\phi}{2}$, $f_{3}=\sin^{2}\frac{\phi}{2}$ and $\Gamma^{\pm}_{h}(\omega)= G_{h}(\omega)n_{h}(\mp\omega)$. $n_{h}(\omega)$ is the average poplulation characterized by the temperature of bath-$h$, $n_{h}(\omega)=(e^{\beta_{h}\omega}-1)^{-1}$ with $\beta_{h}=1/T_{h}$ ($k_B\equiv1$). We assume here the spectral density function of the bath as an Ohmic function with an exponential cutoff: $G_{h}(\omega)=\gamma_{h}\omega e^{-|\omega|/\omega_c}$, one can deduce the bare decay rates $\Gamma^{\pm}_{\mu}$. For example,
\begin{eqnarray}
    \Gamma^{+}_{h}(\omega_{3})&=&\nonumber\textrm{Re}\left[\int_{0}^{\infty }ds {\rm Tr}_{B} [e^{i\omega_{3}s}\sum_{q}\lambda_{q,{h}} b_{q,{h}}e^{-i\omega_{q,{h}}t}\sum_{k}\lambda_{k,{h}}b^{\dag}_{k,{h}}e^{i\omega_{k,{h}}(t-s)}\rho_{B}]\right]\\
   &=&\nonumber \textrm{Re}\left[\int_{0}^{\infty }ds  {\rm Tr}_{B} [ \sum_{q,k}e^{-i(\omega_{q,{h}}-\omega_{k,{h}})t}\lambda_{q,{h}}\lambda_{k,{h}} b_{q,{h}}b^{\dag}_{k,{h}}\rho_{B} e^{-i(\omega_{k,{h}}-\omega_{3})s}] \right]\\
   &=&\nonumber \textrm{Re}\left[\int_{0}^{\infty }ds  \sum_{q}\lambda_{q,h}^2\langle b_{q,{h}}b^{\dag}_{q,{h}}\rangle e^{-i(\omega_{q,{h}}-\omega_{3})s} \right]\\
   &=&\nonumber \sum_{q}\lambda_{q,h}^2 \textrm{Re}\left[\int_{0}^{\infty }ds  e^{-i(\omega_{q,{h}}-\omega_{3})s}\right]\langle b_{q,{h}}b^{\dag}_{q,{h}}\rangle\\
   &=&\nonumber 2\pi\sum_{q}\lambda_{q,h}^2\delta(\omega_{q,h}-\omega_{3})[n_{h}(\omega_{q,h})+1]\\
   &=&\nonumber G_{h}(\omega_{3})[n_{h}(\omega_{3})+1]
\end{eqnarray}
where $n_{h}(\omega)=(e^{\beta_{h}\omega}-1)^{-1}$ with $\beta_{h}=1/T_{h}$ ($k_B\equiv1$) can be defined as the average poplulation characterized by the temperature of bath-$h$.

Collecting all the $12$ terms renders the Redfield master equation. Many time-dependent terms (called the non-secular terms), such as $\tau_{13}\rho_{S}\tau_{13}e^{-2i\omega_{3}t}$, $\tau_{13}\rho_{S}\tau_{12}e^{-i(\omega_{2}+\omega_{3})t}$, and $\tau_{13}\rho_{S}\tau_{21}e^{-i(\omega_{3}-\omega_{2})t}$, then present in Eq.~(\ref{HpH1}).

We denote by $t_{S}$ the timescale of the intrinsic evolution of the system, which is in the same order of a typical value for $|\omega_{m}-\omega_{n}|^{-1}, m\neq n$, involving the energy-spacing of the system. If $t_S$ is much larger than the typical timescale of the relaxation time of the system $t_{R}\sim\gamma_{\mu}^{-1}$, $\mu=h,c,w$, the non-secular terms of rapid oscillation can be safely neglected~\cite{Breuer_book_Open-Quantum_2002}. Thus the condition of performing a full secular approximation is $\gamma_{\mu}\ll|\omega_{m}-\omega_{n}|$.

While in many models as well as ours, although the decay rates set as $\gamma_{h,c,w}/\omega_{a}\sim0.01$ satisfy Born approximation, the energy level $\omega_{b}$ is comparable to $\omega_{a}$ in magnitude. Thus the actual condition is $\omega_{2}+\omega_{3}\gg\omega_{3}-\omega_{2}\sim\gamma_{h,c,w}$. Consequently we have to employ a partial secular approximation by omitting the rapid-oscillating terms, such as $e^{-2i\omega_{3}t}$ and $e^{-i(\omega_{2}+\omega_{3})t}$, yet keeping the slow-oscillating terms, such as $e^{-i(\omega_{3}-\omega_{2})t}$.

Under the partial secular approximation, Eq.~(\ref{HpH1}) is thus rewritten as
\begin{eqnarray}
&& \nonumber{\rm Tr}_{B}\left[H^{h}_{SB}(t)\rho_{S}\rho_{B}H^{h}_{SB}(t-s)\right]
=\Gamma^{+}_{h3}(\omega_{3})\tau_{13}\rho_{S}\tau_{31}
+ \Gamma^{+}_{h1}(\omega_{3})\tau_{13}\rho_{S}\tau_{21}e^{-i(\omega_{3}-\omega_{2})t}
+ \Gamma^{-}_{h3}(\omega_{3})\tau_{31}\rho_{S}\tau_{13}\\
&+&\nonumber\Gamma^{-}_{h1}(\omega_{3})\tau_{31}\rho_{S}\tau_{12}e^{i(\omega_{3}-\omega_{2})t}
+ \Gamma^{+}_{h2}(\omega_{2}) \tau_{12}\rho_{S}\tau_{21}+\Gamma^{+}_{h1}(\omega_{2})\tau_{12}\rho_{S}\tau_{31}e^{-i(\omega_{2}-\omega_{3})t}
+ \Gamma^{-}_{h2}(\omega_{2})\tau_{21}\rho_{S}\tau_{12}\\
&+&\Gamma^{-}_{h1}(\omega_{2})\tau_{21}\rho_{S}\tau_{13}e^{i(\omega_{2}-\omega_{3})t}.\label{HpH2}
\end{eqnarray}
So do all the remaining terms in Eq.~(\ref{HHPP}). Then after rotating back to the Schr\"{o}dinger picture, we attain the Redfield master equation with a partial secular approximation as Eq.~(\ref{ME1}) in the main text.

Following Eq.~(\ref{ME1}), the dynamical equations of the three-level system with inner coupling in terms of matrix elements are given by
\begin{eqnarray}
  \dot{\rho}_{11} &=&\nonumber -2[\Gamma^{-}_{h3}(\omega_{3})+\Gamma^{-}_{h2}(\omega_{2})+\Gamma^{-}_{c2}(\omega_{3})+\Gamma^{-}_{c3}(\omega_{2})]\rho_{11}
  + 2[\Gamma^{+}_{h2}(\omega_{2})+\Gamma^{+}_{c3}(\omega_{2})]\rho_{22}+2[\Gamma^{+}_{h3}(\omega_{3})+\Gamma^{+}_{c2}(\omega_{3})]\rho_{33}\\
  &+&\nonumber  [\Gamma^{+}_{h1}(\omega_{3})-\Gamma^{+}_{c1}(\omega_{3})+\Gamma^{+}_{h1}(\omega_{2})-\Gamma^{+}_{c1}(\omega_{2})](\rho_{23}+\rho_{32}) \\
  \dot{\rho}_{22} &=&\nonumber 2[\Gamma^{-}_{h2}(\omega_{2})+\Gamma^{-}_{c3}(\omega_{2})]\rho_{11}-2[\Gamma^{+}_{h2}(\omega_{2})+\Gamma^{+}_{c3}(\omega_{2})+\Gamma^{-}_{w}(\Omega)]\rho_{22}
  + 2\Gamma^{+}_{w}(\Omega)\rho_{33}-[\Gamma^{+}_{h1}(\omega_{3})-\Gamma^{+}_{c1}(\omega_{3})](\rho_{23}+\rho_{32})  \\
  \dot{\rho}_{33} &=& \nonumber 2[\Gamma^{-}_{h3}(\omega_{3})+\Gamma^{-}_{c2}(\omega_{3})]\rho_{11}-2[\Gamma^{+}_{h3}(\omega_{3})+\Gamma^{+}_{c2}(\omega_{3})+\Gamma^{+}_{w}(\Omega)]\rho_{33}
  +  2\Gamma^{-}_{w}(\Omega)\rho_{22}- [\Gamma^{+}_{h1}(\omega_{2})-\Gamma^{+}_{c1}(\omega_{2})](\rho_{23}+\rho_{32})  \\
  \dot{\rho}_{23}&=&\nonumber [2i\Omega-\Gamma^{+}_{h3}(\omega_{3})-\Gamma^{+}_{c2}(\omega_{3})-\Gamma^{+}_{h2}(\omega_{2})-\Gamma^{+}_{c3}(\omega_{2})-\Gamma^{+}_{w}(\Omega)-\Gamma^{-}_{w}(\Omega)]\rho_{23}
  -[\Gamma^{+}_{h1}(\omega_{3})-\Gamma^{+}_{c1}(\omega_{3})]\rho_{33}
  \\&+&\nonumber [\Gamma^{-}_{h1}(\omega_{3})-\Gamma^{-}_{c1}(\omega_{3})+\Gamma^{-}_{h1}(\omega_{2})-\Gamma^{-}_{c1}(\omega_{2})]\rho_{11}-[\Gamma^{+}_{h1}(\omega_{2})-\Gamma^{+}_{c1}(\omega_{2})]\rho_{22} %
  \\  \dot{\rho}_{32}&=&\nonumber [-2i\Omega-\Gamma^{+}_{h3}(\omega_{3})-\Gamma^{+}_{c2}(\omega_{3})-\Gamma^{+}_{h2}(\omega_{2})-\Gamma^{+}_{c3}(\omega_{2})-\Gamma^{+}_{w}(\Omega)-\Gamma^{-}_{w}(\Omega)]\rho_{32}
  -[\Gamma^{+}_{h1}(\omega_{3})-\Gamma^{+}_{c1}(\omega_{3})]\rho_{33}
  \\  &+&  [\Gamma^{-}_{h1}(\omega_{3})-\Gamma^{-}_{c1}(\omega_{3})+\Gamma^{-}_{h1}(\omega_{2})-\Gamma^{-}_{c1}(\omega_{2})]\rho_{11}-[\Gamma^{+}_{h1}(\omega_{2})-\Gamma^{+}_{c1}(\omega_{2})]\rho_{22}
  .\label{SSequations}
 \end{eqnarray}
As shown by Eq.~(\ref{SSequations}), the off-diagonal (coherence) terms $\rho_{23}$ and $\rho_{32}$ are closely associated with the diagonal terms $\rho_{11}$, $\rho_{22}$ and $\rho_{33}$. Thus the steady state obtained by $\dot{\rho}=0$ may have residue quantum coherence, which is a mark of quantumness.

\section*{References}
\bibliographystyle{elsarticle-num}
\bibliography{quantum-thermal-device}

\end{document}